\documentclass[10pt, letterpaper, conference] {IEEEtran2015}
\usepackage[usenames,dvipsnames]{xcolor}
\usepackage{amsmath}
\usepackage{algorithm}
\usepackage{amssymb}
\usepackage{graphicx}
\usepackage{graphicx,subfigure,url,cite}
\usepackage{pseudocode} 
\usepackage{pgfplots}
\usepackage{tikz}
\usepackage{varwidth}
\usepackage[noend]{algpseudocode}

\newtheorem{property}{Property}

\DeclareGraphicsExtensions{.pdf}

\newcommand{\eat}[1]{}
\newcommand{\ignore}[1]{}

\newcommand{\AMUSE}{\textit{AMuSe}~}
\newcommand{\AMUSENB}{\textit{AMuSe}}
\newcommand{\MUDRA}{\textit{MuDRA}~}
\newcommand{\MUDRANB}{\textit{MuDRA}}
\newcommand{\delete}[1]{{\color{Goldenrod}{\bf Delete:} #1}}

\newcommand{\change}[1]{{\color{black}{}#1}}

\newif\ifTechReport
\TechReportfalse

\newif\ifINFO             
\newif\ifJRNL             

\newif\ifNotINFO       
\newif\ifNotJRNL  

\JRNLtrue

\begin{document}

\title{Experimental Evaluation of Large Scale WiFi Multicast Rate Control}

\author{
Varun Gupta$^\dag$,
Craig Gutterman$^\dag$,
Yigal Bejerano$^\ast$,
Gil Zussman$^\dag$\\ 
{$\dag$ Electrical Engineering, Columbia University, New York, NY, USA.}\\  
 {$\ast$ Bell Labs, Nokia, Murray Hill, NJ, USA}.\\
Email: \{varun@ee, clg2186\}.columbia.edu, bej@research.bell-labs.com , gil@ee.columbia.edu}

\maketitle

\thispagestyle{plain}
\pagestyle{plain}

\begin{abstract}
%
WiFi multicast to \emph{very large groups} has gained attention as a solution for multimedia delivery in crowded areas. Yet, most recently proposed \change{schemes} do not provide performance guarantees and \emph{none have been tested at scale}. \change{To address the issue of providing high multicast throughput with performance guarantees, 
we present the design and experimental evaluation of the Multicast Dynamic Rate Adaptation (\emph{MuDRA}) algorithm.}
\MUDRA balances fast adaptation to channel conditions and stability, which is essential for multimedia applications. \MUDRA relies on feedback from some nodes collected via a light-weight protocol and dynamically adjusts the rate adaptation response time.
\change{Our experimental evaluation of \MUDRA on the ORBIT testbed with over 150 nodes shows that \MUDRA outperforms other schemes and supports high throughput multicast flows to hundreds of receivers while meeting quality requirements. \MUDRA can support multiple high quality video streams, where $90\%$ of the nodes report excellent or very good video quality.}
\end{abstract}



\section{Introduction}
\label{SC:Intro}

Multimedia (e.g., video) delivery is an essential service for wireless networks and several solutions
were proposed for crowded  
\ifJRNL venues~\cite{TYY10,CiscoStadium,YinzCam}\fi
. Most of them are based on dense deployments of Access Points (APs) and require considerable capital and operational expenditure, may suffer from interference between APs, and may exacerbate hidden node 
\ifJRNL problems~\cite{pelechrinis-wintech10,hajlaoui-wintech12}\fi
.
\emph{Multicast} offers another approach for video delivery to large groups of users interested in venue specific content (e.g., sports arenas, entertainment centers, and lecture halls). However, WiFi networks
provide \emph{limited multicast support at a low rate (e.g., $6$Mbps for 802.11a/g) without a feedback mechanism that guarantees service quality}. 
To improve performance, there is a need for a multicast system that \emph{dynamically adapts the transmission rate}~\cite{mcastproblem}. Yet, designing such a system poses several challenges, as outlined below. 

%


\noindent\textbf{Multicast Rate Adaptation (RA) - Challenges:}
\label{SSC:MRAintro}
A key challenge in designing multicast RA schemes for large groups is 
to obtain accurate quality reports with low overhead.
Some \ifJRNL systems~\cite{WMLLM09,CKM+09,Medusa} \fi 
experimentally demonstrated impressive ability to deliver 
video to a few dozen nodes by utilizing  Forward Error Correction (FEC) codes 
and retransmissions. 
However, most approaches do not scale to very large 
groups with hundreds of nodes, due to the following:\\
(i) Most schemes tune the rate to satisfy the receiver with the worst channel condition. As shown in~\cite{amuse,Papagiannaki2006HomeWiFi} 
in crowded venues, a few unpredictable outliers, referred to as {\em abnormal nodes}, may suffer from low SNR and Packet Delivery Ratio (PDR) even at the lowest rate and without interference. This results from effects 
such as multipath and fast fading\ifJRNL~\cite{RF:Rappaport02:wireless_com}\fi.
Therefore,
{\em a multicast scheme cannot provide high rate while ensuring reliable delivery to \underline{all} users.}\\ 
(ii) It is impractical to continuously collect status reports from all or most users without hindering performance. 
Even if feedback is not collected continuously, a swarm of retransmission requests may be sent following an interference event, \ifJRNL(wireless interference is bursty~\cite{Aguayo04interference-locality})~\fi
thereby causing additional interruptions.

\begin{figure}[t]
\raggedleft
\includegraphics[width=0.45\textwidth]{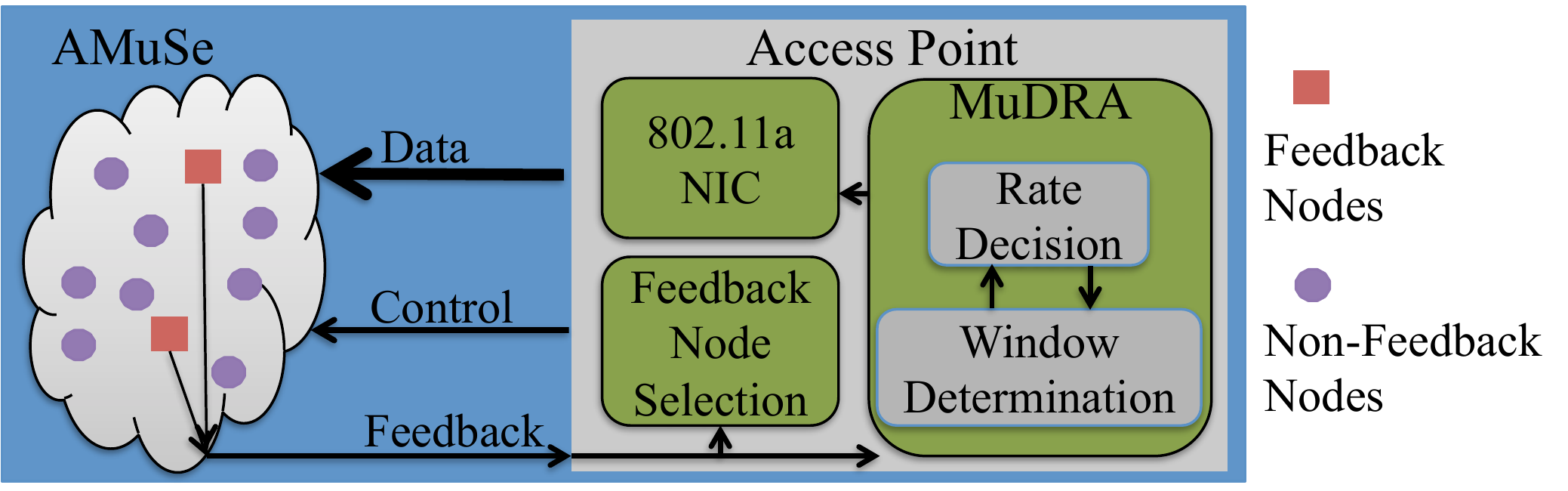}
\caption{The Adaptive Multicast Services (\textit{AMuSe}) system consisting of the Multicast Dynamic Rate Adaptation (MuDRA) algorithm and a multicast feedback mechanism.}
\vspace{-0.2cm}
\label{fig:mudra}
\end{figure}

To overcome these challenges, a multicast system should \emph{conduct efficient RA based on only limited reports from the nodes}. \change{We have been developing the Adaptive Multicast Services (\AMUSENB) system for content delivery over WiFi multicast. In our recent papers~\cite{amuse}, we focused on efficient feedback collection mechanisms for WiFi multicast as part of the \AMUSE system. In this paper, we present the Multicast Dynamic Rate Adaptation (\emph{MuDRA}) algorithm. \MUDRA leverages the efficient multicast feedback collection of \AMUSE and \emph{dynamically adapts the multicast transmission rate to maximize channel utilization while meeting performance requirements}.
Fig.~\ref{fig:mudra} shows the overall \AMUSE system composed of (i) \MUDRA algorithm, and (ii) a feedback mechanism. Before describing \MUDRANB, the overall \AMUSE system design, and our contributions in detail, 
we now first briefly outline the related work relevant to the system design.}

\subsection{Related Work}
\label{SC:RelatedWork}
Unicast RA, multicast feedback schemes, and multicast RA have received considerable 
attention (see \ifJRNL surveys \fi  
\ifJRNL in~\cite{Nguyen2011RAmultiAntSys,Combes2014RAreport,multicastsurvey} \fi
). 

\noindent
{\bf Unicast RA:} We discuss unicast RA schemes, since they can provide insight into the design of multicast RA. 
In \emph{Sampling-based algorithms},
both ACKs after successful transmissions and the relation between the rate and the success probability are used for
RA after several consecutive successful or failed transmissions~\cite{Kamerman97WaveLan,Lacage2004RA,Bicket2005RA}.
The schemes \ifJRNL in~\cite{Pang2005LDARF,Wong2006RRAA,Kim2006CARA} \fi distinguish between losses due to poor channel conditions and collisions, and update the rate based on former.
\ifJRNL
\ifJRNL Recently,~\cite{Radunovic11WhiteSpaces,Combes2014RA} \fi \ifINFO Recently,~\cite{Combes2014RA} \fi propose multi-arm bandit based RA schemes with a statistical bound on the regret.
\fi
However, such schemes 
cannot support multicast, since multicast packets are not acknowledged.
In \emph{Measurement-based schemes} the receiver reports the channel quality to the sender 
which determines the \ifJRNL rate~\cite{Holland2001SNR-RA,Rayanchu2008COLLIE,Judd2008CHARM,Balakrishnan2009SoftRate, Katabi2009FARA,Crepaldi2012CSISF,Belding2013ARAMIS}\fi . 
Most measurement-based schemes modify the wireless driver on the receiver end and some require changes to the standard, which we avoid.

\ifJRNL
\noindent
{\bf Multicast Feedback Mechanisms:}
Solutions for improving multicast service quality are based on collecting feedback from the receivers and adapting the sender rate.
They integrate Automatic Repeat Request (ARQ) mechanisms into the protocol
~\cite{KK01,SHAL02,VCOST07,CKM+09,CSKC10,LH08:BLBP,SR09,WWWZY09}, 
add Forward Error Correction (FEC) packets~\cite{PHKSJ08,AKWP10,CCLKT12,WMLLM09}, and utilize 
RA methods~\cite{SC03,BSS06,VCOST07,LKS12}. 
The feedback mechanisms can be classified into five categories:\\
(i) Collecting {\em Individual Feedback} from all users for each received packet~\cite{SHAL02,WWWG08,SL09,WMLLM09,WWWZY09,FWYL10,802.11aa}. 
Although this offers reliability, it does not scale for large groups. The other approaches provide scalability by compromising on the feedback accuracy.\\
(ii) The {\em Leader-Based Protocol with acknowledgements (LBP-ACK)} method
~\cite{VCOST07,AKWP10,FWYL10,CSKC10,Medusa}
selects a few receivers to provide feedback, 
typically the receivers with the lowest channel quality.\\
(iii) {\em Pseudo-Broadcast}~\cite{CKM+09,PJYK10,Medusa}
converts the multicast feed to a unicast flow and sends it to one leader. 
The leader acknowledges the reception of the unicast flow while the other receivers receive packets by listening to the channel in promiscuous mode.\\
(iv) The {\em Leader-Based Protocol with negative acknowledgements} (LBP-NACK)
~\cite{KK01,LH08:HLBP,LKS12} method improves Pseudo-Broadcast by allowing the other receivers to send NACKs for lost packets.\\ 
%
The leader based approaches (ii)-(iv) cannot provide guarantees on the feedback accuracy~\cite{WWWZY09,amuse}. Moreover, most LBP-ACK and LBP-NACK methods require changes to the standard.\\ 
(v) {\em Cluster-Based Feedback Mechanisms}~\cite{SHAL02,WWWZY09,amuse,amuseGree} handle the scalability issue by using the fact that adjacent receivers experience similar service quality.
They partition the receivers into clusters and select the receiver 
with the weakest channel condition at each cluster as a {\em feedback (FB) node} that sends status reports to the sender. 
These methods, however, do not guarantee reliable delivery to all receivers.\\
Additionally, ~\cite{PHKSJ08,SSFNB09,CCLKT12} propose to use strong FEC for overcoming losses without specifying any feedback mechanism. 
Others~\cite{WWWZY09,CKM+09,Medusa,LKS12} balance between the accuracy requirements and low overhead by using a combination of methods (e.g., Pseudo-Broadcast with infrequent reports from the other receivers).
\fi

\noindent
{\bf Multicast RA:} 
\ifJRNL In~\cite{SC03,BSS06,VCOST07,CKM+09,LSHC11} \fi the sender uses feedback from leaders (nodes with worst channel conditions) for RA. 
In~\cite{LKS12} when the channel conditions are stable, RA is conducted based on reports of a single leader. When the channel conditions are dynamic, feedback is collected from all nodes.
Medusa~\cite{Medusa} combines Pseudo-Multicast with infrequent application layer feedback reports from all nodes.
The MAC layer feedback sets backoff parameters while application layer feedback is used for RA and retransmissions of video packets.
\change{Recently, in ~\cite{amuse} we considered multicast 
to a large set of nodes and provided a rudimentary RA scheme which is not designed to achieve optimal rate, maintain stability, or respond to interference.}

\subsection{Our Contributions} 
\label{SSC:Contribution}


\change{
We present a multicast rate adaptation algorithm \MUDRA which is designed to support WiFi multicast to hundreds of users in crowded venues.
\MUDRA can provide high throughput while ensuring high \emph{Quality of Experience} (QoE). \MUDRA benefits from a large user population, which allows selecting a small yet sufficient number of Feedback (FB) nodes with marginal channel conditions for monitoring the  quality. We address several design challenges related to appropriate configuration of the feedback level. 
We note that using \MUDRA does not require any modifications to the IEEE 802.11 standard or the mobile devices. 
We implemented \MUDRA with the \AMUSE system on the ORBIT testbed~\cite{ORBIT}, evaluated its performance with all the operational IEEE 802.11 nodes (between 150--200), and compared it to other multicast schemes.} Our key contributions are:

\noindent
{\bf (i) The need for RA:} We empirically  
demonstrate the importance of RA. Our experiments on ORBIT show that when the multicast rate exceeds an optimal rate, termed as \emph{target-rate}, numerous receivers suffer from low PDR and their losses cannot be recovered. 
We also observed that even a controlled environment, such as ORBIT, can 
suffer from significant interference.
These observations constitute the need for a stable and interference agnostic RA algorithm that {\em does not exceed the target-rate}. 

\noindent
{\bf (ii) Practical method to detect the target-rate:}
Pseudo-multicast schemes that rely on unicast RA \cite{CKM+09} may occasionally sample higher rates and retreat to a lower rate after a few failures.
Based on the observation above about the target rate, schemes with such sampling mechanisms will provide low QoE to many users. 
To overcome this, we developed a method to detect when the system operates at the target-rate, termed the {\bf target condition}.
Although the target condition is sufficient but not necessary, our experiments show that it is almost always satisfied when transmitting at the target-rate.
\MUDRA makes RA decisions based on the target condition and  employs a dynamic window based mechanism to avoid rate changes due to small interference bursts.

\noindent
{\bf (iii) Extensive experiments with hundreds of receivers:}
Our experiments demonstrate that \MUDRA swiftly converges to the target-rate, while meeting the Service Level Agreement (SLA) requirements (e.g., ensuring PDR above $85\%$ to at least $95\%$ of the nodes).
Losses can be recovered by using appropriate application-level FEC 
\ifJRNL methods~\cite{PHKSJ08,SSFNB09,CCLKT12,Nybomand07FEC,Ababneh14FEC}\fi
.

\MUDRA is experimentally compared to (i) pseudo-multicast with a unicast RA~\cite{minstrel}, (ii) fixed rate, and (iii) a rate adaptation mechanism proposed in~\cite{amuse} which we refer to as Simple Rate Adaptation (SRA) algorithm. \MUDRA achieves 2x higher throughput than pseudo-multicast while sacrificing PDR only at a few poorly performing nodes. 
While the fixed rate and SRA schemes can obtain similar throughput as \MUDRANB, they do not meet the SLA requirements. Unlike other schemes, \MUDRA preserves high throughput even in the presence of interference. Additionally, \MUDRA can handle significant node mobility.
Finally, we devise a live multicast video delivery approach for \MUDRANB. We show that in our experimental settings with target rate of $24-36$Mbps, \MUDRA can deliver 3 or 4 high definition H.264 videos (each one of $4$Mbps) where over $90\%$ of the nodes receive video quality that is classified as excellent or good based on user perception.

\change{To summarize, to the best of our knowledge, \MUDRA is the first multicast RA algorithm designed to satisfy the specific needs of multimedia/video distribution in crowded venues. Moreover, \AMUSE in conjunction with \MUDRA is the first multicast content delivery system that has been evaluated \emph{at scale}.}
\ifJRNL
The rest of the paper is organized as follows. Section~\ref{SC:ORBIT} describes the ORBIT testbed and important observations. Section~\ref{SC:Model} presents the model and objectives.
\MUDRANB's design is described in Sections ~\ref{SC:Algorithm} and ~\ref{SC:ReportingRate}. The experimental evaluation is presented in Section~\ref{SC:Results} before concluding in Section~\ref{SC:Conclusion}.
\fi

\section{Testbed and Key Observations}
\label{SC:ORBIT}

We evaluate \MUDRA on the ORBIT testbed~\cite{ORBIT}, which is a dynamically configurable grid of $20\times 20$ (400) 802.11 nodes where the separation between nodes is $1$m. 
It is a good environment to evaluate \MUDRANB, since it provides a very large and dense population of wireless nodes, 
similar to the anticipated crowded venues.
%

\noindent
{\bf Experiments:}
To avoid performance variability due to a mismatch of WiFi hardware and software, only nodes equipped with {\em Atheros 5212/5213} 
cards with {\em ath5k} driver were selected. 
For each experiment {\em we activated \underline{all} the operational nodes} that meet these specifications (between $150$ and $250$ nodes). 
In all the experiments, one corner node served as a single multicast AP. The other nodes were multicast receivers.
The AP used 
802.11a to send a multicast UDP flow, where each packet was 1400 bytes.
{\em The AP used the lowest supported transmission power of $1$mW $=$ $0$dBm 
to ensure that the channel conditions of some nodes are marginal.}

\begin{figure}
    \centering
    \includegraphics[trim=8mm 0mm 8mm 0mm, width=0.4\textwidth]{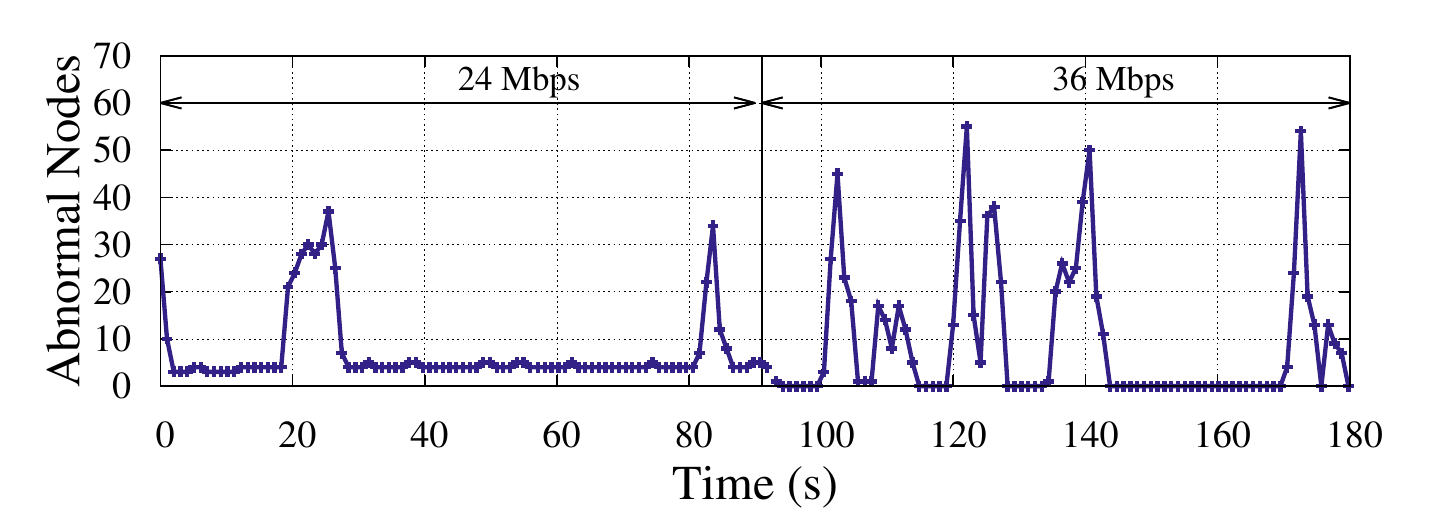}
   \caption{Experimental measurement of the number of abnormal nodes in time, for fixed rates of $24$ and $36$Mbps.}
   \label{FIG:fixedrate}
\end{figure}

\begin{figure}
    \centering
     \vspace*{-0.3cm}
    \includegraphics[trim=8mm 0mm 8mm 0mm, width=0.4\textwidth]{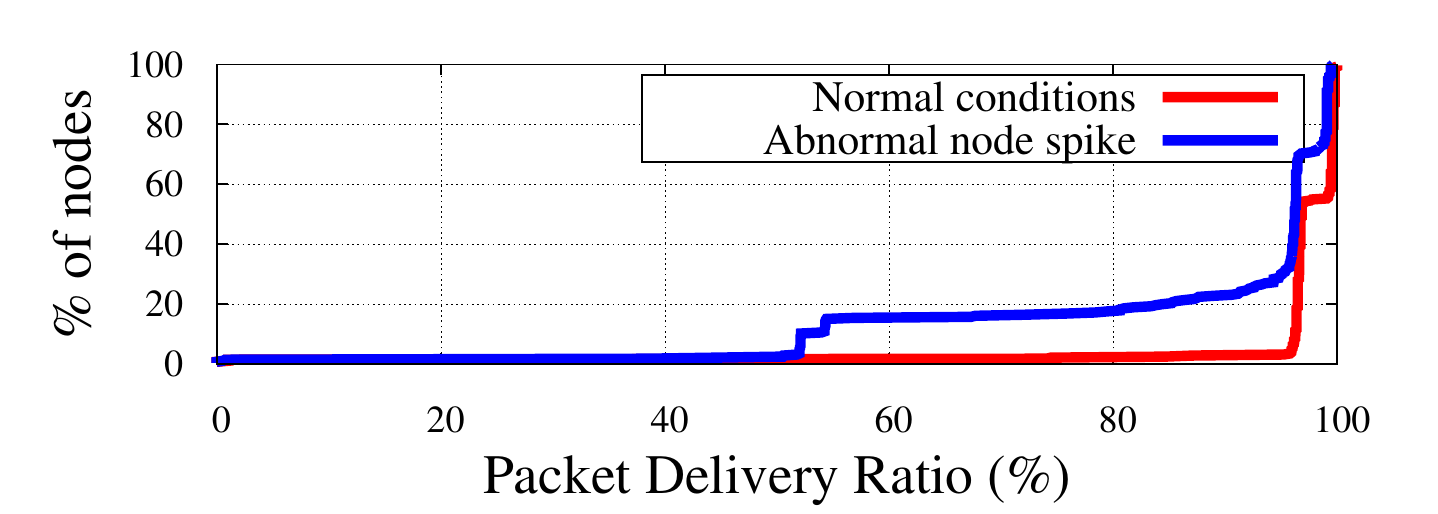}
   \caption{The CDF of the PDR values of $170$ nodes during normal operation and during a spike at rate of $36$Mbps.}
   \label{FIG:pdrcdf36M}
   \vspace*{-0.2cm}
\end{figure}

\ifJRNL
\noindent
{\bf Technical challenges:} 
While analyzing the performance, we noticed that clients disconnect from the AP at high bit-rates, thereby causing performance degradation. This results from the fact that increasing the bit-rate also increases the WiFi beacon bit-rate which may not be decoded at some nodes. A sustained loss of beacons leads to node disconnection. To counter this, we modified the ath5k driver to send beacons at the minimum bit-rate.
\fi

\noindent
{\bf Interference and Stability:}
We study the time variability of the channel conditions on the ORBIT testbed by measuring the number of nodes with low PDR (below a threshold of $85\%$).
We call these nodes \emph{abnormal nodes} (the term will be formally defined in Section~\ref{SC:Model}). The number of abnormal nodes out of 170 nodes for rates of $24$ and $36$Mbps is shown in Fig.~\ref{FIG:fixedrate}.
We repeated these experiments several times and observed that even at a low rate, the channel may suffer from sporadic interference events, which cause a sharp increase in the number of abnormal nodes. These interference spikes caused by non-WiFi devices are beyond our control and their duration varies in time.

Fig.~\ref{FIG:pdrcdf36M} provides the Cumulative Distribution Function (CDF) of the PDR values with and without sporadic interference.
The figure shows that during a spike, over $15\%$ of the nodes suffer from PDR around $50\%$. Further, the location of the nodes affected by the spikes varies with time and does not follow a known pattern.
These experiments show that even in a seemingly controlled environment, {\em nodes may suffer from sporadic continuous interference, 
which may cause multicast rate fluctuations}.
Users are very sensitive to changes in video \ifJRNL quality~\cite{cranley2006user, balachandran2012quest}, \fi and therefore, to keep a high QoE we would like to avoid rate changes due to sporadic interference. 

\ifJRNL
\begin{table}[t]
\caption{Notation and parameter values used in experiments.}
\vspace{0.3cm}
{\footnotesize
\centering
\begin{tabular}{|p{0.9cm}|p{5.6cm}|p{1.1cm}|}\hline
{\bf Symbol} & ~~~~~~~~~~~~~~~~~~~~~ {\bf Semantics} & {\bf Exp. Val.}\\ \hline
$n$ & Number of nodes associated with the AP. & $>150$\\ \hline
$X$ & {\em Population threshold} - Minimal fraction of nodes that should experience high PDR. & $95\%$\\ \hline
$A_{max}$ & The maximal number of allowed abnormal nodes.& $8$\\ \hline
$L$ & {\em PDR threshold} - Threshold between acceptable (normal) and low (abnormal) PDR. & $85\%$ \\ \hline
$H$ & Threshold between high PDR and mid-PDR. & $97\%$ \\ \hline
$K$ & Expected number of FB nodes, $K=\alpha\cdot A_{max}$. & $30$\\ \hline
$R$ & Reporting PDR threshold. &\\ \hline
$A_t$  & Number of abnormal nodes at time $t$. &\\ \hline
$M_t$  & Number of mid-PDR FB nodes at time $t$. &\\ \hline
$W_{min}$ &  Minimal RA window size (multiples of reporting intervals). & $8$\\ \hline
$W_{max}$ &  Maximal RA window size. & $32$\\ \hline
\end{tabular}
}
\label{TAB:notation}
\vspace{-0.2cm}
\end{table}
\fi

\section{Network Model and Objective}
\label{SC:Model}


We consider a WiFi LAN with multiple APs and frequency planning such that the
transmissions of adjacent APs do not interfere with each other.
Thus, {\em for RA we consider 
a single AP with $n$ associated users}. We assume low mobility (e.g., users watching a sports event).
Although we consider a controlled environment, 
the network may still suffer from sporadic 
interference, as shown in Section~\ref{SC:ORBIT}.
The main notation used in the paper is summarized in Table~\ref{TAB:notation}.
Specifically, a {\em PDR-Threshold} $L$, is defined such that a node has high QoE if its PDR is above $L$.  
Such a node is called a {\em normal node}. Otherwise, it is considered an {\em abnormal node}.

Our \emph{objective is to develop a practical and efficient rate control system} which satisfies the following requirements:
\\
{\bf (R1) High throughput --} Operate at the highest possible rate, i.e., the {\em target rate}, while preserving SLAs.\\
{\bf (R2) Service Level Agreements (SLAs) --} Given $L$ (e.g., $L=85\%$), and a {\em Population-Threshold} $X$ (e.g., $X=95\%$), 
the selected rate should guarantee that at least $X\%$ of the nodes experience PDR above $L$ (i.e., are normal nodes).
Except for short transition periods, this provides an upper bound of $A_{max} = \lceil n\cdot(1-X)\rceil$ on the 
number of permitted abnormal nodes.\\
{\bf (R3) Scalability --} Support hundreds of nodes.\\
{\bf (R4) Stability --} Avoid rate changes due to sporadic channel condition changes.\\
{\bf (R5) Fast Convergence --} Converge fast to the target rate after long-lasting changes (e.g., user mobility or network changes).\\
{\bf (R6) Standard and Technology Compliance --} 
No change to the IEEE 802.11 standard or operating system of the nodes.

\section{Multicast Rate Adaptation}
\label{SC:Algorithm}

\noindent
\change{The overall multicast rate adaptation process of \MUDRA as a part of the \AMUSE system relies on three main components, as illustrated in Fig.~\ref{fig:mudra} and discussed below.} We first provide a high level description of each component and then discuss the details in the following subsections.

\noindent
{\bf (i) Feedback (FB) Node Selection:} Selects a small set of {\em FB nodes} that provide reports for making RA decisions. 
We describe the FB node selection process in Section~\ref{SSC:FBM} and calculate the reporting interval duration in Section~\ref{SC:ReportingRate}.\footnote{Unlike in unicast where each packet is acknowledged, \MUDRANB's reporting intervals are long (in the experiments we consider 2 reports per second).}

The following two components compose the \MUDRA Algorithm (Algorithm~\ref{ALG:rateadaptationprocess}). It collects the PDR values from the FB nodes, updates their status (normal or abnormal), invokes the \textsc{GetRate} procedure,
which calculates the desired rate, and invokes the \textsc{GetWinSize} procedure, which determines the window size of rate updates (to maintain stability).

\noindent
{\bf (ii) Rate Decision} (Procedure \ref{ALG:rateadaptation}): Utilizes the limited and infrequent FB reports to determine the highest possible rate, termed the {\em 
target-rate}, while meeting the requirements in Section~\ref{SC:Model}. 
The rate decisions (lines 5--15) rely on rate decision rules that are described in  Section~\ref{SSC:RateSelection}.
To maintain rate stability, rate change operations are permitted, only if the conditions for rate change are satisfied for time equal to a window size (determined by the \emph{Stability Preserving Method}). 

\begin{algorithm}[t]
\caption{\MUDRA Algorithm}
\label{ALG:rateadaptationprocess}
\begin{algorithmic}[1]    
{\footnotesize
\State $rate \gets lowestRate$, $window \gets W_{min}$, $changeTime \gets t$, $refTime \gets t, t := \textrm{current time}$
\While {($true$)}
  \State Get PDR reports from all FB nodes
  \State Get Status of each FB node $i$
  \State Calc $\hat{A}_{t}$ and $\hat{M}_{t}$
  \State $rate, action, changeTime \gets GetRate(...)$
  \State $window, refTime \gets GetWinSize(...)$
  \State set multicast rate to $rate$
  \State sleep one reporting interval 
\EndWhile
}
\end{algorithmic}
\end{algorithm}

\begin{algorithm}[t]
\floatname{algorithm}{Procedure}
\setcounter{algorithm}{0}
\caption{Rate Decision}
\label{ALG:rateadaptation}
\begin{algorithmic}[1]
{\footnotesize
\Procedure{GetRate}{$rate, window, changeTime ,t$}
\State $action \gets Hold$
\If {$(t-changeTime) > window $}
  \State $canDecrease \gets true$, $canIncrease \gets true$
  \For{$\tau \gets 0 \TO window$}
     \If{$\hat{A}_{t-\tau}<A_{max}$}
        \State $canDecrease \gets  false$
     \ElsIf {$\hat{A}_{t-\tau} + \hat{M}_{t-\tau} > A_{max}$}
        \State $canIncrease \gets  false$ 
     \EndIf  
  \EndFor 
  \If{$canDecrease$ and $rate > rate_{min}$}
      \State $rate \gets NextLowerRate$
      \State $action \gets Decrease$, $changeTime \gets t$   
  \EndIf
  \If{$canIncrease$ and $rate < rate_{max}$}
      \State $rate \gets NextHigherRate$
      \State $action \gets Increase$, $changeTime \gets t$
   \EndIf
\EndIf\\
\Return{$rate$, $action$, $changeTime$}
\EndProcedure
}
\end{algorithmic}
\end{algorithm}

\begin{algorithm}[t]
\floatname{algorithm}{Procedure}
\caption{Window Size Determination}
\label{ALG:window}
\begin{algorithmic}[1]  
{\footnotesize
\Procedure{GetWinSize}{$Action, window, refTime, t$}
 \If {$Action = Decrease$} 
   \State $window \gets \min(W_{max},2 \cdot window)$, $refTime \gets t$
 \ElsIf {$Action = Increase$}
   \State $refTime \gets t$
 \ElsIf { $(t - refTime) > thresholdTime$ \\
~~~~~~~~and $Action = Hold$}
   \State $window \gets \max(W_{min}, window - 1)$
   \State $refTime \gets t$
\EndIf\\
\Return{$window, refTime$}
\EndProcedure
}
\end{algorithmic}
\end{algorithm}

\noindent
{\bf (iii) Stability Preserving Method} (Procedure~\ref{ALG:window}): A window based method that maintains rate stability 
in the event of sporadic interference and after an RA decision.
It follows the classical Additive Increase Multiplicative Decrease (AIMD) approach.
The duration of the time window varies according to the network and channel characteristics (e.g., the typical duration of interference).
More details appear in Section~\ref{SSC:window}. 

\subsection{Feedback Node Selection}
\label{SSC:FBM}

\MUDRA uses a simple and efficient mechanism based on 
a quasi-distributed FB node selection process, 
termed {\em $K$-Worst}~\cite{amuse}, where the 
AP sets the number of FB nodes and their reporting rates.
$K$ nodes with 
the worst channel conditions are selected as FB nodes (the node's channel condition is determined by its PDR).
Hence, the selection process ensures an upper bound 
on the number of FB messages, regardless of the multicast group size.
This upper bound is required for limiting the interference from FB reports, as explained in Section~\ref{SC:ReportingRate}.
\ifJRNL
The process works as follows:
At the beginning of each reporting interval the AP sends a message 
with a list of $K$ or less FB nodes as well as a 
{\em reporting PDR threshold} $R$. 
$R$ is used for adjusting the set of FB nodes to changes due to 
mobility or variation of the channel condition, i.e., interference%
\footnote{when the system is activated the FB list is empty and $R=L$.}.
Upon receiving this message, each FB node waits a short random time for avoiding collisions and then reports its measured PDR to the AP.
Every other node checks if its PDR value is below $R$ and in such 
situation it volunteers to serve as an FB node.
To avoid a swarm of volunteering messages in the case of sporadic 
interference, a non FB node verifies that its PDR values are below $R$ for three consecutive reporting intervals before volunteering.
At the end of a reporting interval, the AP checks the PDR values 
of all the FB and volunteering nodes, it selects the $K$ with lowest 
PDR values as FB nodes and updates $R$.
If the number of selected FB nodes is $K$ then for keeping the 
stability of the FB list, $R$ is set slightly below (e.g., $1\%$) 
the highest PDR value of the FB nodes.
Otherwise, $R$ is set slightly (e.g., $0.5\%$) above the highest 
PDR value of the FB nodes.
The AP sends a new message and the process repeats.
\fi

\begin{figure}[t]
\centering
\includegraphics[width=0.4\textwidth]{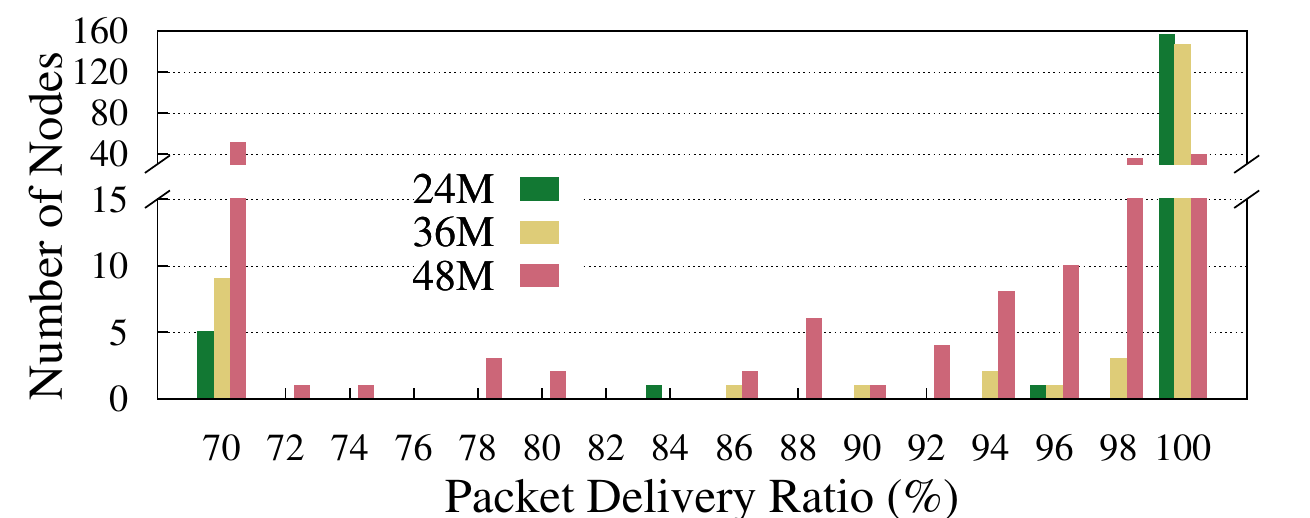}
\caption{The PDR distribution of one set of experiments with $TX_{AP}$ rates of $24$, $36$, and $48$Mbps.
}
\vspace*{-0.2cm}
\label{FIG:PDRdist}
\end{figure}

\begin{figure}[t]
\centering
\includegraphics[width=0.4\textwidth]{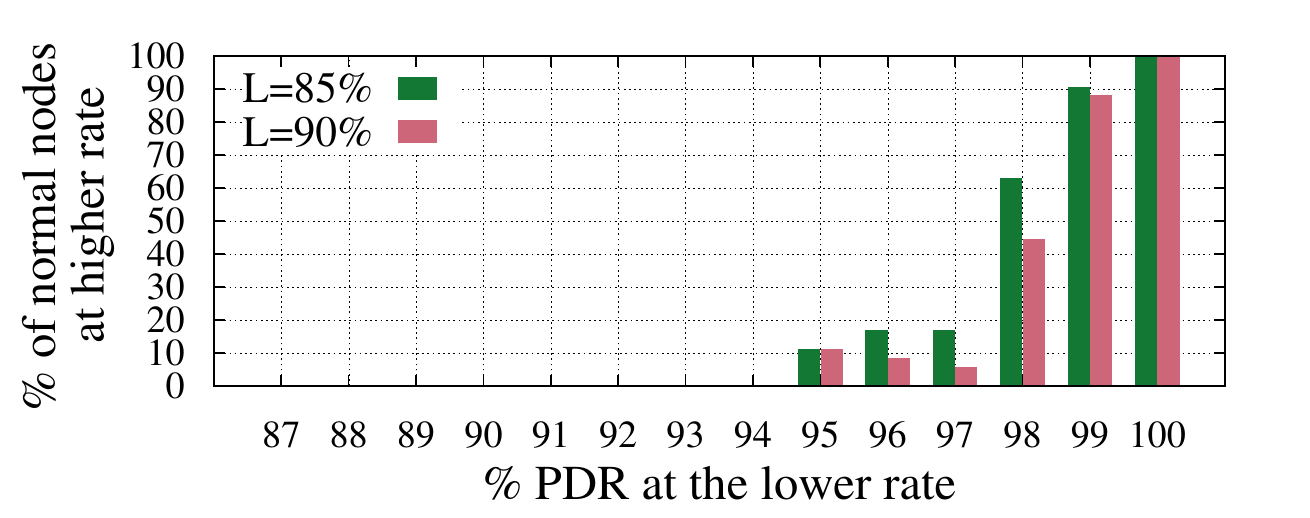}
\caption{The percentage of nodes that remain normal after increasing the $TX_{AP}$ from $36$Mbps to $48$Mbps vs.~their PDR values at the $36$Mbps for different PDR-thresholds ($L$).}
\label{FIG:offline}
\vspace*{-0.2cm}
\end{figure}

\subsection{Rate Decision Rules and Procedure}
\label{SSC:RateSelection}

In this subsection, we describe the {\em target condition} which is an essential component of the rate selection rules. Then, we describe the rules and the corresponding Procedure~\ref{ALG:rateadaptation}.

\noindent
{\bf The Target Condition:}
At a given time, the FB reports are available only for the current rate.
To detect the target-rate, most RA schemes occasionally sample higher rates. 
However, the following experiment shows that this approach may cause undesired disruption to many receivers.
We evaluated the PDR distribution of $160-170$ 
nodes for different multicast transmission rates, denoted as $TX_{AP}$ for 3 different experiment runs on different days.
Fig.~\ref{FIG:PDRdist} shows the number of nodes in different PDR ranges for $TX_{AP}$ values of $24$, $36$, and $48$Mbps for one experiment with $168$ nodes.
When $TX_{AP}$ is at most $36$Mbps, the number of abnormal nodes is very small (at most 5).
However, when $TX_{AP}$ exceeds $36$Mbps, the PDR of many nodes drops significantly.
In this experiment $47$ nodes became abnormal nodes which is more than $A_{max}=8$ (for $X=95\%$).
We observed similar results in other experiments.
Thus, in this case, the {\em target rate} is $36$Mbps which is the highest rate above which the SLA requirements will be violated. We observed similar results for other experiments as well.

A key challenge is to {\em determine if the AP operates at the target-rate, without FB reports from higher rates.}
We refer to this assessment as the {\em target condition}. 
Unfortunately, the target-rate cannot be detected from RF measurements, such as SNR. As shown in~\cite{Reis2006interference, Halperin2010wifi} 
different nodes may have different receiver sensitivities, which may result in substantial PDR gaps between nodes with similar RF measurements. However, large scale multicast environments enable us to efficiently predict the target condition as described next.

From Fig.~\ref{FIG:PDRdist}, we obtain the following important observation.

\noindent
\emph{Observation I}: When operating below the target-rate, almost all the nodes have PDR close to $100\%$.
However, when operating at the target-rate, noticeable number of receivers experience PDR below $97\%$. 
At $36$Mbps, $17$ nodes had PDR below $97\%$, 
which is substantially more than $A_{max}=8$.

Fig.~\ref{FIG:offline} shows the average percentage of nodes that remain normal vs.~their initial PDR when increasing $TX_{AP}$ from $36$Mbps to $48$Mbps averaged for 3 different sets of experiments. The total number of nodes in these experiments was $168$. We derive the following observation from Fig.~\ref{FIG:offline}.

\noindent
\emph{Observation II}: There is a PDR threshold, $H=97\%$, 
such that every node with PDR between $L$ and $H$ becomes abnormal after the rate increase with very high probability. Note that $97\%$ is the highest threshold for which this observation holds. 
We refer to these nodes as {\em mid-PDR nodes}.

Observation II is not surprising. As reported in~\cite{Reis2006interference,souryal},
each receiver has an SNR band
of $2-5$dB, in which its PDR drops from almost $100\%$ to almost $0\%$. 
The SNR of mid-PDR nodes lies in this band.
Increasing the rate requires $2-3$dB higher SNR at the nodes. 
Hence, mid-PDR nodes with SNR in the transition band before the rate increase will be below or at the lower end of the transition band after the increase, and therefore, become abnormal nodes. 

In summary, Observations I and II imply that it is possible to assess the target condition by monitoring the nodes close to transitioning from normal to abnormal. Let $A_t$ and $M_t$ denote the number of abnormal and mid-PDR nodes 
at time $t$, respectively.
We obtain the following empirical property. \begin{property}[Target Condition]
\label{PR:optCond}
Assume that at a given time $t$, the following condition holds,
\begin{equation}
\label{EQ:targetCond}
A_t\leq A_{max}~~~and~~~A_t+ M_t > A_{max}
\end{equation}
then almost surely, the AP transmits on the target-rate at time $t$. This is sufficient but not a necessary condition.
\end{property}

It is challenging to analytically predict when the target condition is satisfied with the available FB information and without a model of the receiver sensitivity
of all nodes. However, our experiments show that the target condition is typically valid when operating at the target-rate.

\noindent
{\bf Adjusting the Multicast Rate:}
The SLA requirement (R2) and target condition (\ref{EQ:targetCond}) give us a clear criteria for changing the rate.
The FB scheme only gives us estimates of $A_t$ and $M_t$, denoted by $\hat{A}_t$ and $\hat{M}_t$ respectively. 
For the $K$-Worst scheme, if $K > A_{max} + \delta$ ($\delta$ is a small constant), then $\hat{A}_t$ and $\hat{M}_t$ are sufficient to verify if (\ref{EQ:targetCond}) is satisfied because of the following property:

\begin{property}
\label{PR:St}
If $K \geq A_{max} + \epsilon$, then, $\hat{A}_t = \min(A_t,A_{max}+\epsilon)$ and 
$\hat{A}_t+\hat{M}_t = min(A_t+M_t, A_{max}+ \epsilon)$, where $\hat{A}_t$ and $\hat{M}_t$ are the known number of abnormal and mid-PDR known to the AP, and $\epsilon$ is a small constant. In other words, given that $K$ is large enough, the $K$-worst scheme provides accurate estimates of abnormal and mid-PDR nodes.
\end{property}

\ifJRNL
\noindent
\emph{Proof:}
First consider the proof for $\hat{A}_t = \min(A_t,A_{max}+\epsilon)$. If $A_t \leq A_{max}+\epsilon$, then $\hat{A}_t = A_t$ since $K \geq A_{max} + \epsilon$ and all abnormal nodes must belong in the K FB nodes set. If $A_t > A_{max}+\epsilon$ then all the FB nodes chosen are abnormal and $\hat{A}_t = A_{max}+\epsilon$. 
A similar argument can be made for establishing $\hat{A}_t+\hat{M}_t = min(A_t+M_t, A_{max}+ \epsilon)$. If $A_t+M_t \leq A_{max}+ \epsilon$, then $\hat{A}_t+\hat{M}_t = A_t+M_t$. If $A_t+M_t > A_{max}+ \epsilon$, then $\hat{A}_t+\hat{M}_t$ is upper bounded by $A_{max}+\epsilon$.
\fi

The objective is to choose minimum $K$ (for minimum FB overhead) that is sufficient to verify (\ref{EQ:targetCond}).
In our experiments, we found that for $A_{max} = 8$, $K > 10$ works well (Section~\ref{SSC:static}).
We now derive the following {\em rate changing rules}:

\noindent 
\textbf{Rule I} $\hat{A}_t > A_{max}$: The system violates 
the SLA requirement (R2) and the rate is reduced.

\noindent 
\textbf{Rule II} $\hat{A}_t+\hat{M}_t\geq A_{max}-\epsilon$: The system satisfies the target condition.

\noindent 
\textbf{Rule III} $\hat{A}_t+\hat{M}_t < A_{max}-\epsilon$: 
The target condition does not hold and the rate can be increased,
under the stability constraints provided in Section~\ref{SSC:window}.

\noindent In our experiments we use $\epsilon=2$ to prevent rate oscillations.

The rate change actions in Procedure~\ref{ALG:rateadaptation} are based on the these rules.
The flags $canIncrease$  and $canDecrease$ indicate whether the multicast rate should be increased or decreased.
Rate change operations are permitted only if the time elapsed since the last rate change is larger than the window size determined by the \emph{Stability Preserving Method} (line 3).
The for-loop checks whether the rate should be decreased according to Rule I (line 6) or increased according to Rule II (line 9) for the window duration.
Finally, based on the value of the flags and the current rate, the algorithm determines the rate change operation and updates the parameters $rate$ and $action$, accordingly (lines 10--15).


\subsection{The Stability Preserving Method}
\label{SSC:window}

\ifJRNL
\begin{figure}[t]
\centering
\includegraphics[scale=0.5]{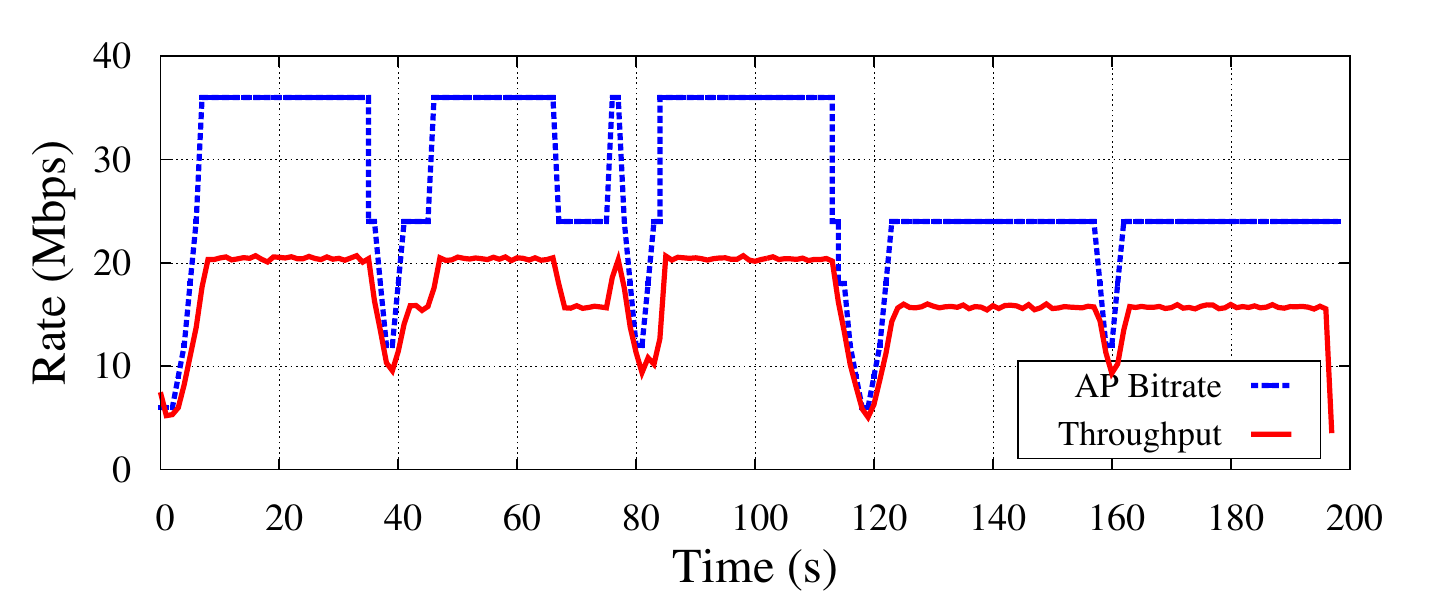}
\caption{Evolution of the multicast rate over time when the delay between rate changes = $1$s ($2$ reporting intervals).}. 
\label{FIG:instable}
\vspace{-0.4cm}
\end{figure}
\fi

It is desirable to change the rate as soon as Rules I or III are satisfied to minimize QoE disruption (see (R5) in Section~\ref{SC:Model}).
\ifJRNL
However, as we show in Fig.~\ref{FIG:instable} such a strategy 
may cause severe fluctuations of the transmission rate. 
\fi
These result from two main reasons:
(i) the reporting mechanism not stabilizing after the last rate change, and
(ii) interference causing numerous low PDR reports. 

To address this, we introduce in Procedure~\ref{ALG:window} a {\em window based RA technique} which considers the two situations and balances fast convergence with stability. 
In Procedure~\ref{ALG:rateadaptationprocess}, rate is changed only if the rate change conditions are satisfied over a given {\em time window}, after the last rate change operation (lines 5-9). 
To prevent oscillations due to short-term wireless channel degradation, when the rate is reduced,
the window is doubled in Procedure~\ref{ALG:window} (line 3).
%
The window size is decreased by $1$ when a duration $thresholdTime$ elapses from the last rate or window size  change (line 8). This allows recalibrating the window  after an atypical long interference episode.
The window duration varies between $W_{min}$ and $W_{max}$ FB reporting periods. In the experiments, $W_{min}=8$ and $W_{max}=32$. 

\subsection{Handling Losses}
\label{SSC:Losses}

\MUDRA can handle mild losses (below $15\%$) by adding 
application level FEC~\cite{Nybomand07FEC} to 
the multicast streams.
The PDR-Threshold in our experiments ($L=85\%$) was selected to allow nodes to handle 
losses in the event of short simultaneous transmission of another node. 
In such a situation, the collision probability is below 
$2/CW_{min}$, where $CW_{min}$ is the minimal 802.11 contention window. 
For 802.11a/g/n $CW_{min}=16$, which implies collision  
probability is below $12.5\%$.
Therefore, nodes with high PDR (near $100\%$) should be able to 
compensate for the lost packets. 
If there is strong interference, other means should be used.
For instance, the multicast content can be divided into high and 
low priority flows, augmenting the high priority flow with 
stronger FEC during the interference period, while postponing  
low priority flows. 


\begin{figure*}[t]
\centering
\subfigure[]{
\includegraphics[trim=5mm 0mm 5mm 0mm, width=0.3\textwidth]{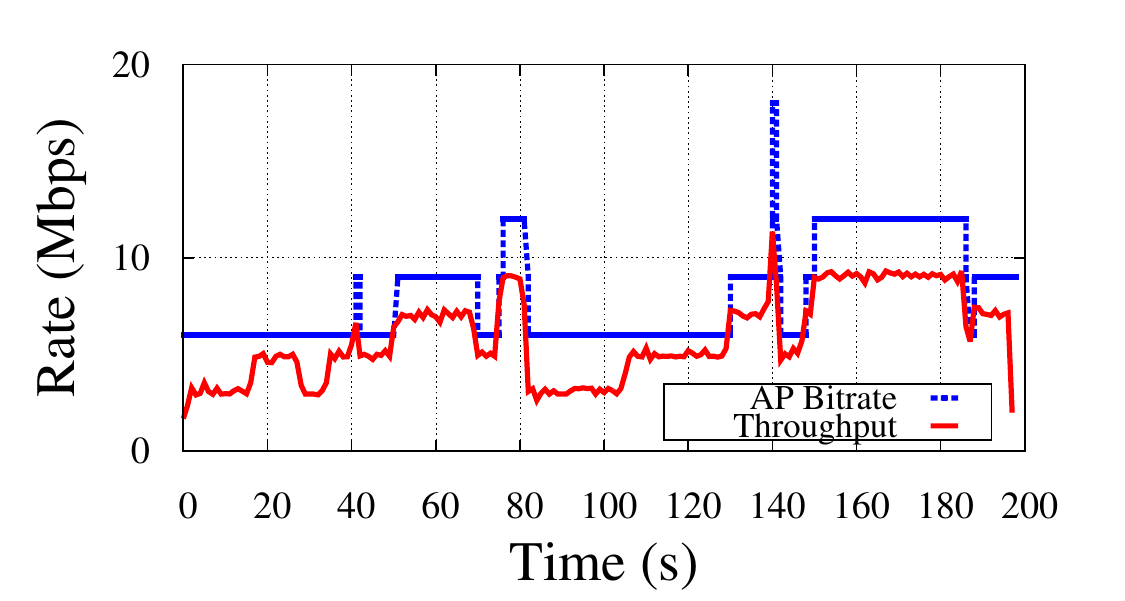}
\label{FIG:reportingexample} 
}
\subfigure[]{
\includegraphics[trim=5mm 0mm 5mm 0mm, width=0.3\textwidth]{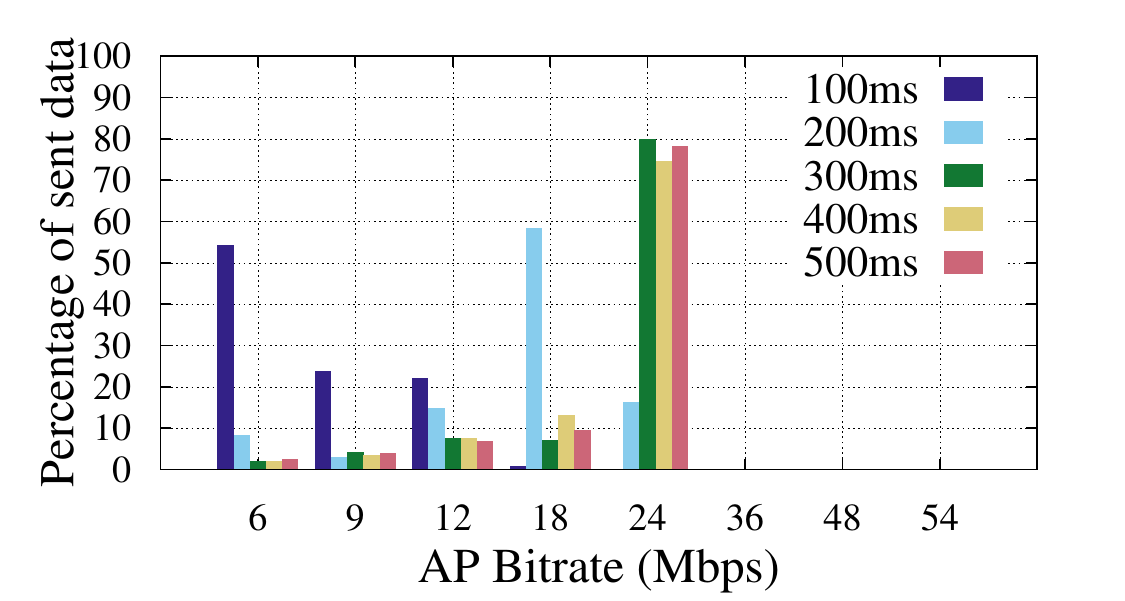}
\label{FIG:reportingthroughput} 
}%
\subfigure[]{
\includegraphics[trim=5mm 0mm 5mm 0mm, width=0.3\textwidth]{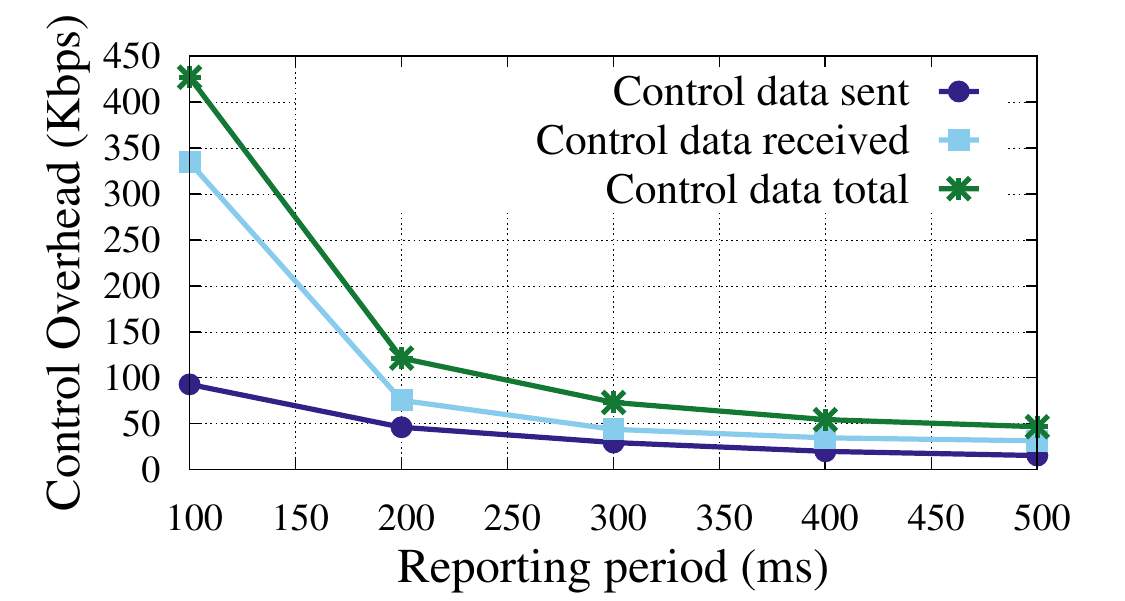} 
\label{FIG:reportingcontrol} 
}
\caption{(a) Rate adaptation performance for reporting intervals of 100ms, (b) Fraction of data sent at various rates with \MUDRA for different reporting intervals, and (c) Control overhead for various reporting intervals.}
\vspace{-0.4cm}
\label{FIG:Exp1}
\end{figure*}

\section{Reporting Interval Duration}
\label{SC:ReportingRate}

\MUDRA relies on status reports from the FB nodes.
For immediate response to changes in service quality, the status reports should be sent as frequently as possible, 
(i.e., minimal {\em reporting interval}).
However, this significantly impairs the system performance as described below.

\noindent
{\bf Impact of Aggressive Reporting:} 
%
Figs.~\ref{FIG:reportingexample}-\ref{FIG:reportingcontrol} show the impact of different reporting intervals on \MUDRA.
In these experiments, the number of FB nodes ($K$) is 50 and the total number of nodes is 158.
To focus on RA aspects, we set both $W_{min}$ and $W_{max}$ to 5 reporting intervals. 
Fig.~\ref{FIG:reportingexample} shows that when the reporting interval is too short, \MUDRA does not converge to the target rate of $24$Mbps.
Fig.~\ref{FIG:reportingthroughput} shows that in the case of reporting interval of $100$ms, more than 50\% of the packets are transmitted at the lowest rate of $6$ Mbps. Fig.~\ref{FIG:reportingcontrol} shows that the control overhead is significantly larger for short reporting intervals (shorter than $200$ms). The control overhead comprises of unicast FB data sent by nodes and multicast data sent by AP to manage $K$ FB nodes.

These phenomena result from collisions between feedback reports and multicast messages. In the event of a collision, FB reports, which are unicast messages, are retransmitted, while multicast messages are lost.
Frequent reporting increases the collision probability, 
resulting in PDR reduction and causes the classification of many 
nodes as mid-PDR nodes, i.e., $PDR<H_{high}=97\%$.
Thus, due to Rule II  from Section~\ref{SSC:RateSelection}, the rate is 
kept close to the minimal rate.

\noindent
{\bf Appropriate Reporting Interval Duration:}
Assume a greedy AP which continuously transmits multicast messages.
We now estimate the PDR reduction, denoted as $\Delta{PDR}$,
for a given reporting interval $T$ and upper bound $K$ 
on the number of FB nodes (both normal and abnormal), 
when the system operates at the low rate of $6$ Mbps.
In this rate, the transmission duration of multicast and 
FB messages are $D\approx 3.0$ms and $d\approx 1$ms. 

\begin{table}
  \caption{The percentage of $\Delta{PDR}(T)$}
  \centering
  {\footnotesize
  \begin{tabular}{| l | c | c| c| c | c| c| c|} \hline
   T (ms)        & 100  & 200  & 300  & 400  & 500  & 700 & 1000   \\ \hline
   $\Delta{PDR}\%$& 4.69 & 1.56 & 0.94 & 0.67 & 0.52 & 0.36 & 0.25 \\ \hline
  \end{tabular}
  }
  \label{TB:DPDR}
\vspace*{0cm}
\end{table}


\ifJRNL 
Assume a greedy AP which continuously transmits multicast messages.
We now estimate the PDR reduction, denoted as $\Delta{PDR}$,
for a given reporting interval $T$ and upper bound $K$ 
on the number of FB nodes (both normal and abnormal), 
when the system operates at the low rate of $6$Mbps.

\noindent
{\bf Packet Transmission Duration:}
We denote with $D$ and $d$ the transmission duration of multicast and feedback report message at the rate of $6$Mbps, respectively.
Since the length of each multicast packet is $12 Kbits$, its 
transmission duration is $\frac{12 \text{Kbits}}{6 \text{Mbps}} = 2.0$ms. 
Given WiFi overhead of about $30\%$, we assume $D=3$ms. 
The feedback messages are much shorter and we assume that their 
transmission duration is $d=1$ms.

\noindent
{\bf Number of feedback reports and multicast messages:}
Consider a time interval $U$, say a minute. 
The number of feedback reports, denoted as $F$, is\\ 
$~~~~~~~~~~~~~~~~~~~~~~~~~~~~~~~~F=\frac{U}{T}\cdot K$\\
The number of multicast message $B$ is given by,
\[
B=\frac{U-d\cdot F}{D}=\frac{U}{D}\cdot\left(1-\frac{d\cdot K}{T}\right)
\]

\noindent
{\bf Collision probably of a multicast packet ($\Delta{PDR}$):}
Let us first calculate the number of contention window slots, denoted by $S$, in which packet may be transmitted from the view point of the AP
during the time interval $U$. 
Recall that between any two multicast transmissions, the AP waits 
an average of half of the contention window size $CW_{min}/2=8$.
This leads to\\
$~~~~~~~~~~~~~~~~~~~~~~~~~~~~~~~~ S=\frac{CW_{min}}{2}\cdot B$

$\Delta{PDR}$ is the fraction of slots in which both the AP and a FB node send a message. 
To simplify our estimation, we ignore collisions and retransmission of FB messages%
\footnote{These are second order effects of already low collision probabilities.},
and assume that in any slots only one FB node may transmit.
Therefore,
\[
\Delta{PDR}=\frac{F}{S}\cdot\frac{B}{S}=
\left[\frac{2}{CW_{min}}\right]^2\cdot\frac{F}{B}
\]
With proper assignment we get, 
\begin{equation}
\label{EQ:DPDR}
\Delta{PDR}=\left[\frac{2}{CW_{min}}\right]^2\cdot\frac{K\cdot D}{T-d\cdot K}
\end{equation}

\begin{table}
  \caption{The percentage of $\Delta{PDR}(T)$}
  \centering
  \begin{tabular}{| l | c | c| c| c | c| c| c|} \hline
   T (ms)        & 100  & 200  & 300  & 400  & 500  & 700 & 1000   \\ \hline
   $\Delta{PDR}\%$& 4.69 & 1.56 & 0.94 & 0.67 & 0.52 & 0.36 & 0.25 \\ \hline
  \end{tabular}
  \label{TB:DPDR}
  \vspace*{-0.2cm}
\end{table}

Equation (\ref{EQ:DPDR}) confirms that $\Delta{PDR}$ is reduced by 
increasing the reporting interval or by using a higher bit-rate, 
which reduces $D$.
Table~\ref{TB:DPDR} provides the $\Delta{PDR}$ values for $K=50$ when $T$ varies between $0.1$ to $1$ second. 
In our experiments we wanted $\Delta{PDR} \leq 0.5\%$, 
which implies using reporting interval $T\ge 500$ms.
\fi 

\section{Experimental Evaluation} 
\label{SC:Results}

\begin{figure*}[t]
\centering
\subfigure[]{
\includegraphics[trim=5mm 0mm 5mm 0mm, width=0.3\textwidth]{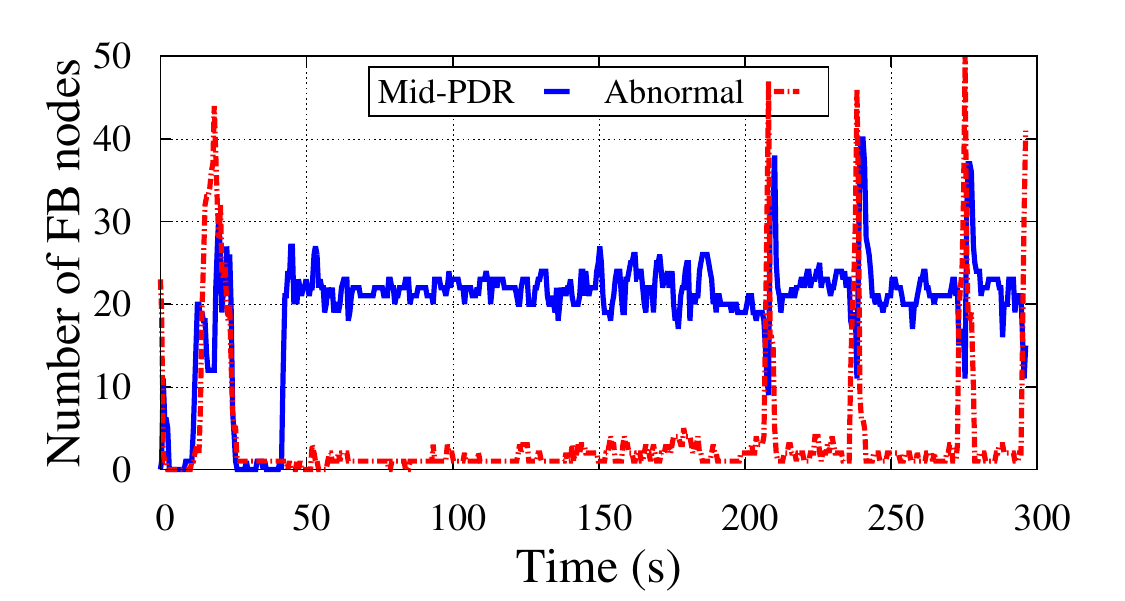}
\label{FIG:abnormaladaptivewindow} 
}%
\subfigure[]{
\includegraphics[trim=5mm 0mm 5mm 0mm, width=0.3\textwidth]{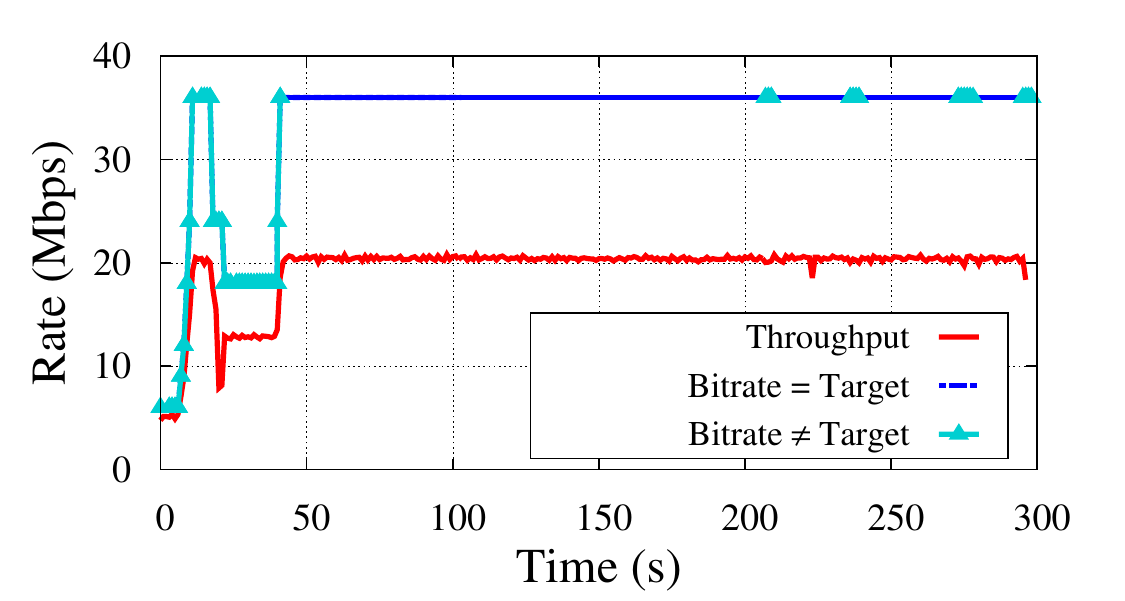} 
\label{FIG:rateadaptivewindow} 
}
\subfigure[]{
\includegraphics[trim=5mm 0mm 5mm 0mm, width=0.3\textwidth]{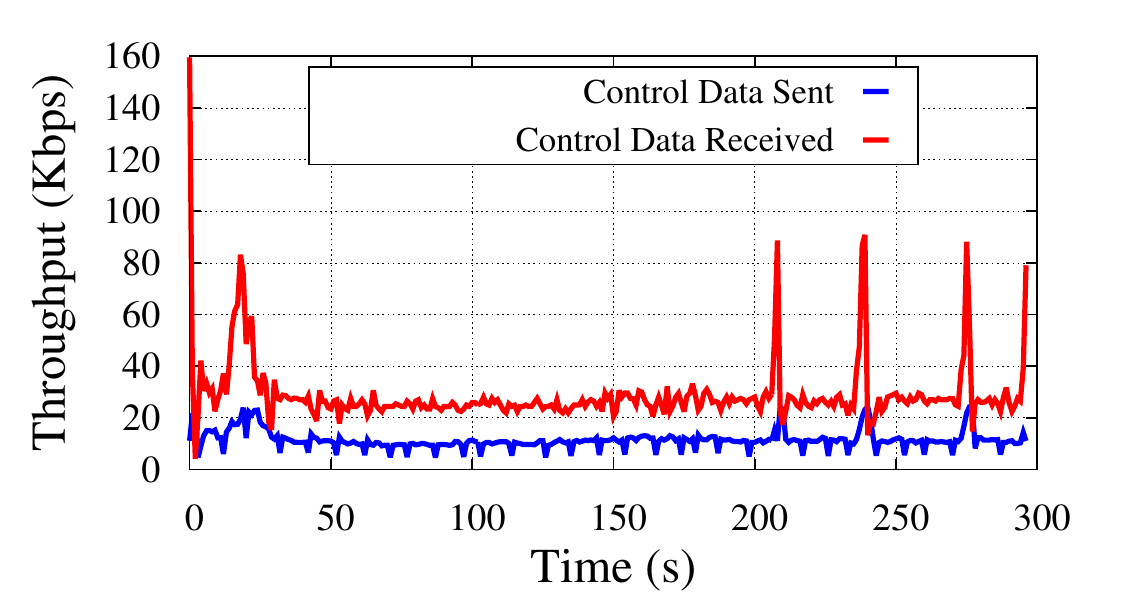}
\label{FIG:controladaptivewindow} 
}
\caption{A typical sample of \MUDRANB's operation over $300$s with 162 nodes: (a) Mid-PDR and abnormal nodes, (b) Multicast rate and throughput measured at the AP, and (c) Control data sent and received.}
\vspace{-0.4cm}
\label{FIG:Final}
\end{figure*}

\begin{figure*}[t]
\centering
\subfigure[]{
\includegraphics[trim=5mm 2mm 5mm 7mm, width=0.3\textwidth]{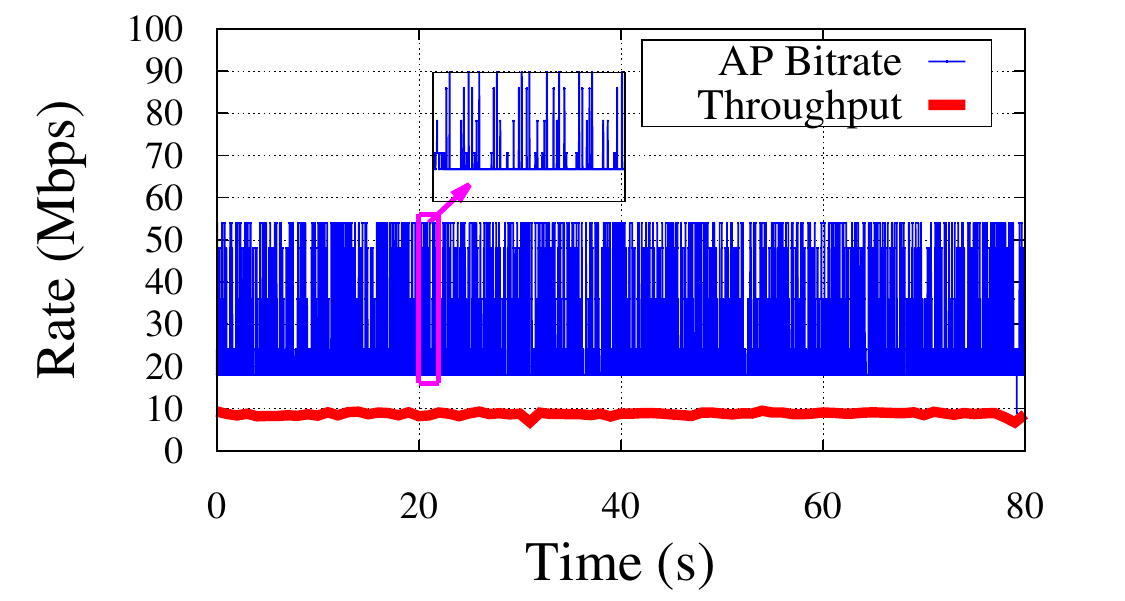}
\label{FIG:PMthroughput} 
}
\subfigure[]{
\includegraphics[trim=5mm 0mm 5mm 0mm, width=0.3\textwidth]{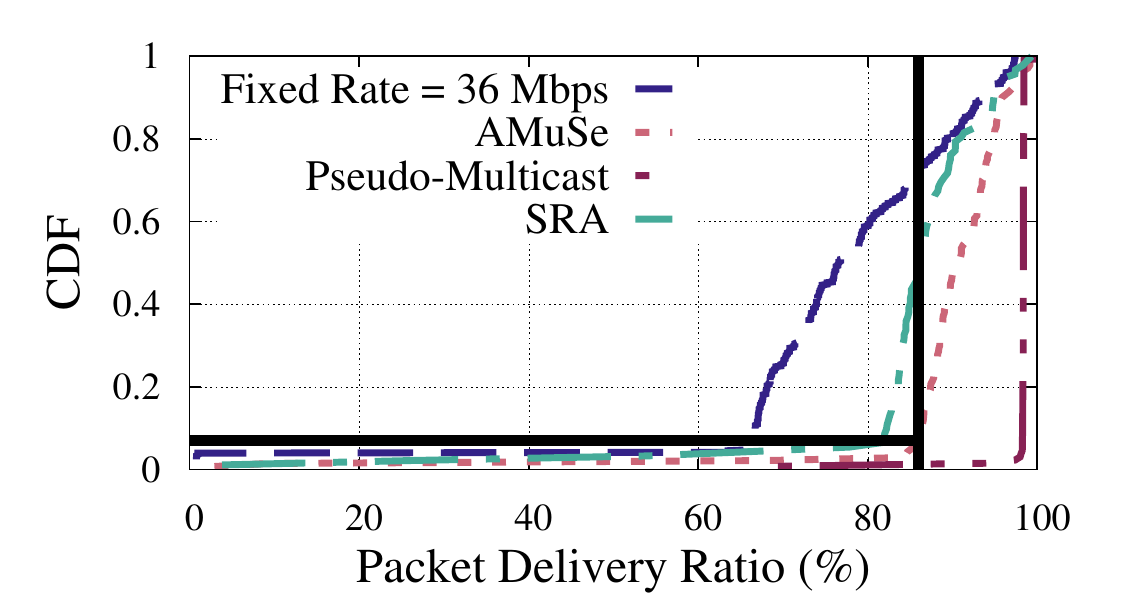}
\label{FIG:RAComparison} 
}%
\subfigure[]{
\includegraphics[trim=5mm 0mm 5mm 0mm, width=0.3\textwidth]{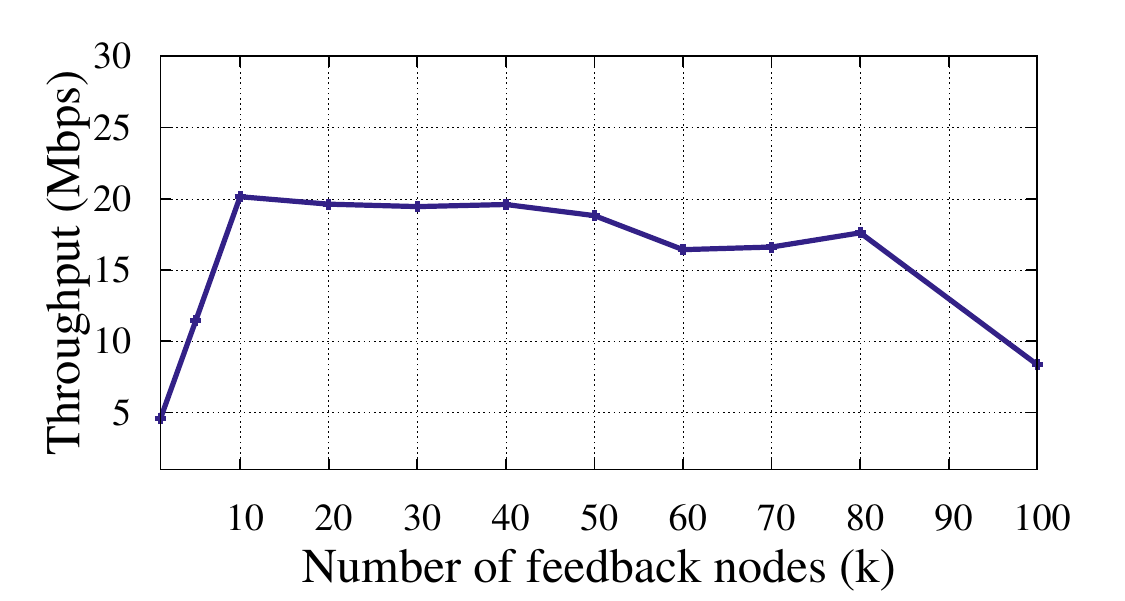} 
\label{FIG:FBComparison} 
}
\caption{(a) Rate and throughput for the pseudo-multicast scheme, (b) CDF of PDR distributions of 162 nodes for fixed rate, \MUDRA, Pseudo-Multicast, and SRA schemes, and (c) Multicast throughput vs.\ the number of feedback nodes ($K$).}
\vspace*{-0.2cm}
\label{FIG:Comparison}
\end{figure*}


For evaluating the performance of \MUDRA on
the ORBIT testbed, we use the parameter values listed in Table~\ref{TAB:notation}.
%
The performance metrics are described below:\\ 
(i) {\em Multicast rate and throughput}: The time instants when the target condition is satisfied are marked separately.\\
(ii) {\em PDR at nodes}: Measured at each node.\\
(iii) {\em Number of abnormal and mid-PDR nodes}: We monitored {\em all} the abnormal and mid-PDR nodes (not just the FB nodes).\\
(iv) {\em Control traffic}: The feedback overhead (this overhead is very low and is measured in Kbps).

We compared \MUDRA to the following schemes:\\
(i) {\em Fixed rate scheme}: Transmit at a fixed rate of $36$Mbps, since it is expected to be the target rate.\\
(ii) {\em Pseudo-multicast}: Unicast transmissions to the node with the lowest SNR/RSS. The unicast RA is the driver specific RA algorithm \emph{Minstrel} \cite{minstrel}. The remaining nodes are configured in promiscuous mode. \\
(iii) {\em Simple Rate Adaptation (SRA) algorithm \cite{amuse}}: This scheme also relies on measuring the number of abnormal nodes for making RA decisions. Yet, it is not designed to achieve the target rate, maintain stability, or respond to interference.

\subsection{Performance Comparison}
\label{SSC:static}
We evaluated the performance of \MUDRA in several experiments on different day with $160-170$ nodes. Fig.~\ref{FIG:Final} shows one instance of such an experiment over $300$s with $162$ nodes.
Fig.~\ref{FIG:abnormaladaptivewindow} shows the mid-PDR and abnormal nodes for the duration of one experiment run. Fig.~\ref{FIG:rateadaptivewindow} shows the rate determined by \MUDRA. The AP converges to the target rate after the initial interference spike in abnormal nodes at $15$s.
The AP successfully ignored the interference spikes at time instants of $210$, $240$, and $280$s to maintain a stable rate.
The target-condition is satisfied except during the spikes.
The overall control overhead as seen in Fig.~\ref{FIG:controladaptivewindow} is approximately $40$Kbps. The population of abnormal nodes stays around $2-3$ for most of the time which implies that more than $160$ nodes ($>98\%$) have a PDR $>85\%$. The actual throughput is stable at around $20$Mbps which after accounting for $15\%$ FEC correction implies a goodput of $17$Mbps. 

Fig.~\ref{FIG:PMthroughput} shows a sample of the throughput and rate performance of the pseudo-multicast scheme. The throughput achieved is close to $9$Mbps. We observe that pseudo-multicast frequently samples higher rates (up to $54$Mbps) leading to packet losses.
The average throughput for different schemes over 3 experiments of $300$s each (conducted on different days) with $162$ nodes is shown in Table~\ref{TB:RAThroughput}.
\MUDRA achieves 2x throughput than pseudo-multicast scheme. The fixed rate scheme yields approximately $10\%$ higher throughput than \MUDRANB. SRA has similar throughput as \MUDRA.

\begin{table}
  \caption{Average throughput (Mbps) of pseudo-multicast, \MUDRA, and SRA schemes with and without background traffic.}
  \vspace*{0.2cm}
  \centering
  {\footnotesize
  \begin{tabular}{| l | c | c | }
  \hline
    & No Background traffic & Background traffic
   \\ \hline
   Fixed rate = $36$Mbps & 20.42 & 13.38 \\ \hline
   Pseudo-Multicast & 9.13 & 5.36 \\ \hline
   \MUDRA & 18.75 & 11.67 \\ \hline
   SRA & 19.30 & 4.55 \\ \hline
  \end{tabular}
  }
  \label{TB:RAThroughput}
  \vspace*{-0.3cm}
\end{table}

Fig.~\ref{FIG:RAComparison} shows the distribution of average PDR of 162 nodes for the same 3 experiments. 
In the pseudo-multicast scheme, more than $95\%$ of nodes obtain a PDR close to $100\%$ (we did not consider any retransmissions to nodes listening in promiscuous mode). \MUDRA meets the QoS requirements of $95\%$ nodes with at least $85\%$ PDR.
On the other hand, in SRA and the fixed rate schemes $45\%$ and $70\%$ of the nodes have PDR less than $85\%$, respectively. 

In pseudo-multicast, more reliable transmissions take place at the cost of reduced throughput, since the AP communicates with the node with the poorest channel quality in unicast.
The significant difference in QoS performance of the fixed rate and SRA schemes is because the target rate can change due to interference, etc.
 In such a situation, \MUDRA can achieve the new target rate while the fixed rate and SRA schemes lead to significant losses (we observed that exceeding the target rate even $10\%$ of time may cause up to $20\%$ losses and less than $5\%$ throughput gain).

\noindent
\textbf{Changing number of FB nodes:} We varied the number of FB nodes ($K$) between $1-100$ for \MUDRANB. Fig.~\ref{FIG:FBComparison} shows the throughput as $K$ changes. For $K = 1$, \MUDRA tunes to the node with the worst channel quality, and consequently, the throughput is very low. 
On the other hand, increasing $K$ from $30$ to $90$ adds similar amount of FB overhead as decreasing the report interval from $500$ms to $200$ms in Section~\ref{SC:ReportingRate}. Thus, the throughput decreases for a large number of FB nodes. The throughput for $K$ between $10-50$ does not vary significantly which is aligned with our discussion in Section~\ref{SC:Algorithm} that \MUDRA needs only $K > A_{max} + \delta$ for small $\delta$ to evaluate the target rate conditions.

\ifJRNL
\begin{figure}[t]
\centering
\includegraphics[trim=5mm 0mm 10mm 0mm, width=0.3\textwidth]{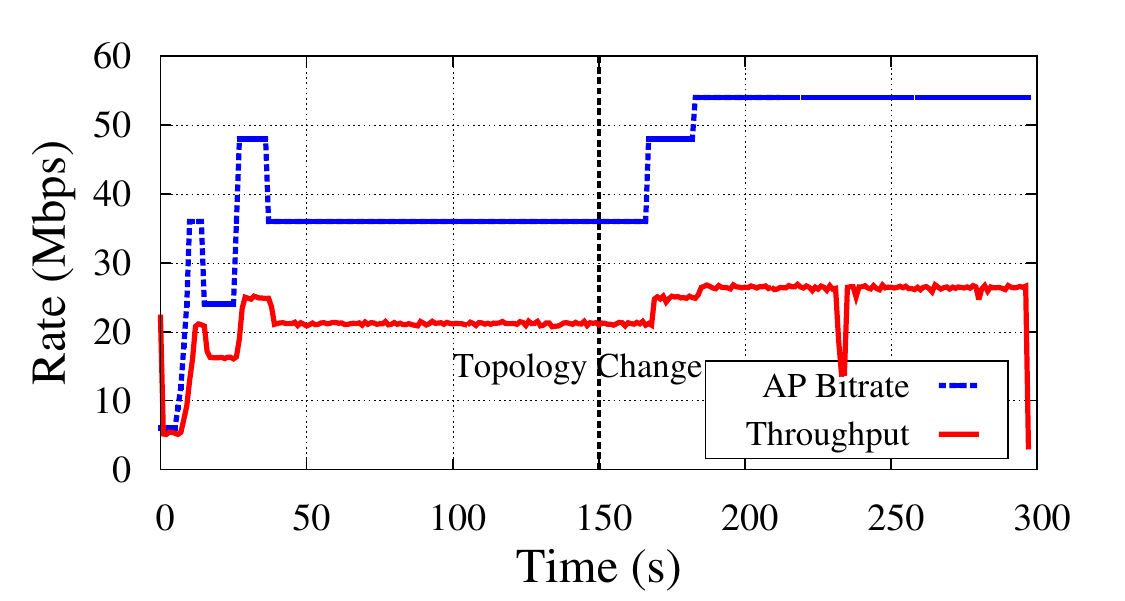}
\caption{Emulating topology change by turning off FB nodes after $150$s results in changing optimal rate for \MUDRA.}
\label{FIG:blacklist}
\vspace{-0.25cm}
\end{figure}

\noindent
\textbf{Impact of topology changes}: To demonstrate that changes in the network may lead \MUDRA to converge to a different rate, we devised a strategy to emulate network topology changes on the grid. During an experiment, a number of FB nodes are turned off at a given time.
Since FB nodes have the lowest PDRs, it may lead to changes in the target rate as a large number of nodes with low PDR disappear from the network. Fig.~\ref{FIG:blacklist} shows the scenario when $30$ FB nodes are turned off after $150$s during the experiment. The rate converges quickly and without oscillations to a new target rate of $54$Mbps. 
\fi

\subsection{Impact of Mobility}
\label{SSC:Mobility}

\begin{figure*}[t]
\centering
\subfigure[]{
\includegraphics[trim=5mm 0mm 5mm 0mm, width=0.3\textwidth]{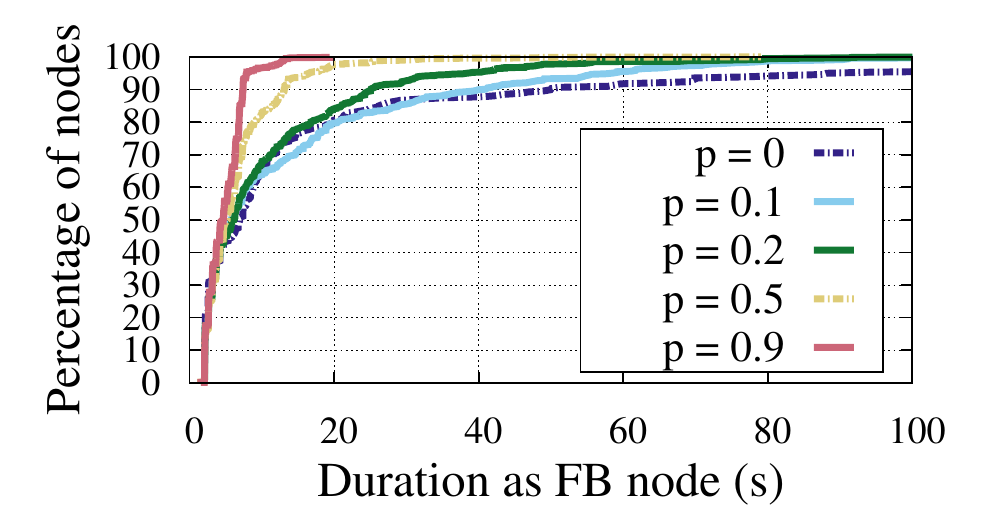}
\label{FIG:mobilityage} 
}%
\subfigure[]{
\includegraphics[trim=5mm 0mm 5mm 0mm, width=0.3\textwidth]{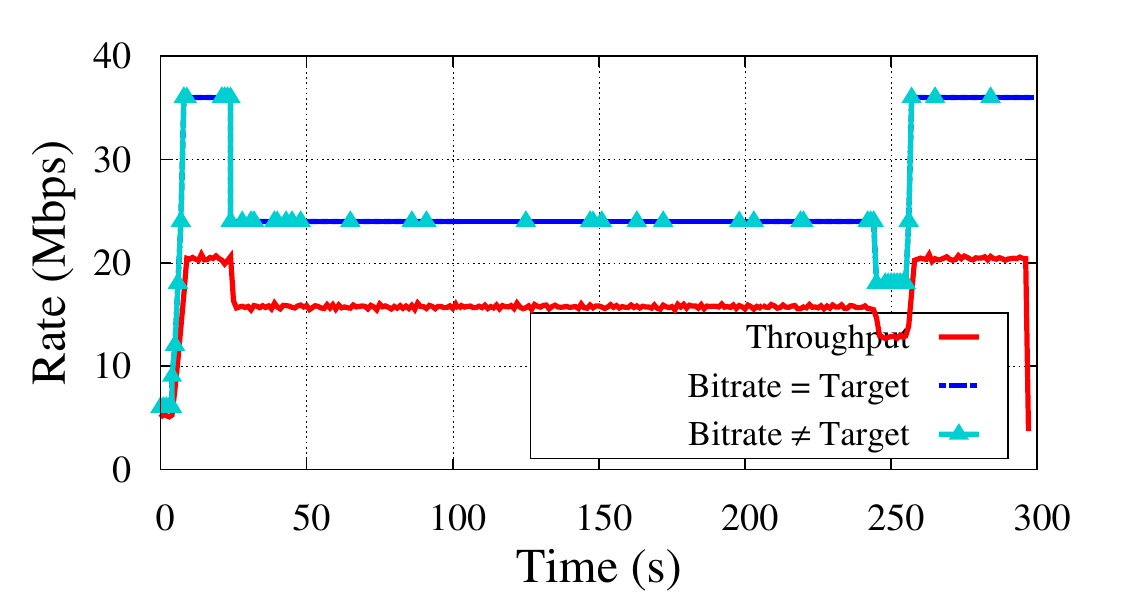}
\label{FIG:ratep=20} 
}%
\subfigure[]{
\includegraphics[trim=5mm 0mm 5mm 0mm, width=0.3\textwidth]{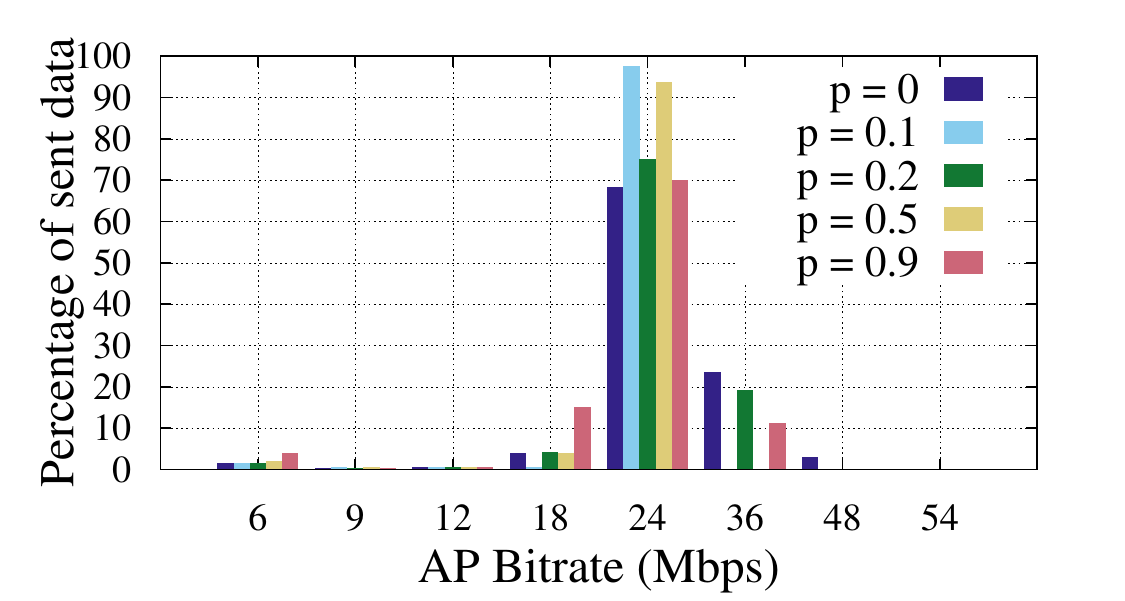}
\label{FIG:mobilityrate} 
}
\caption{Performance of \MUDRA with emulated node mobility: (a) Distribution of time durations for which a node is a feedback node for different values of probability $p$ of node switching its state on/off every $6$s, (b) RA with $p = 0.2$, (c) Percentage of data sent at various rates for different values of $p$.}
\vspace*{-0.4cm}
\label{FIG:mobility}
\end{figure*}

We evaluate \MUDRA performance when emulating severe mobility conditions.
In the experiments, each node leaves or joins the network with probability $p$ after every $6$s. Thus, $p=0.1$ implies that a node changes its state with probability of approximately $50$\% at least once in a minute.
Initially, $50$\% of the nodes are randomly selected to be in the network.

\begin{figure*}[t]
\centering
\subfigure[]{
\includegraphics[trim=5mm 0mm 5mm 0mm, width=0.3\textwidth]{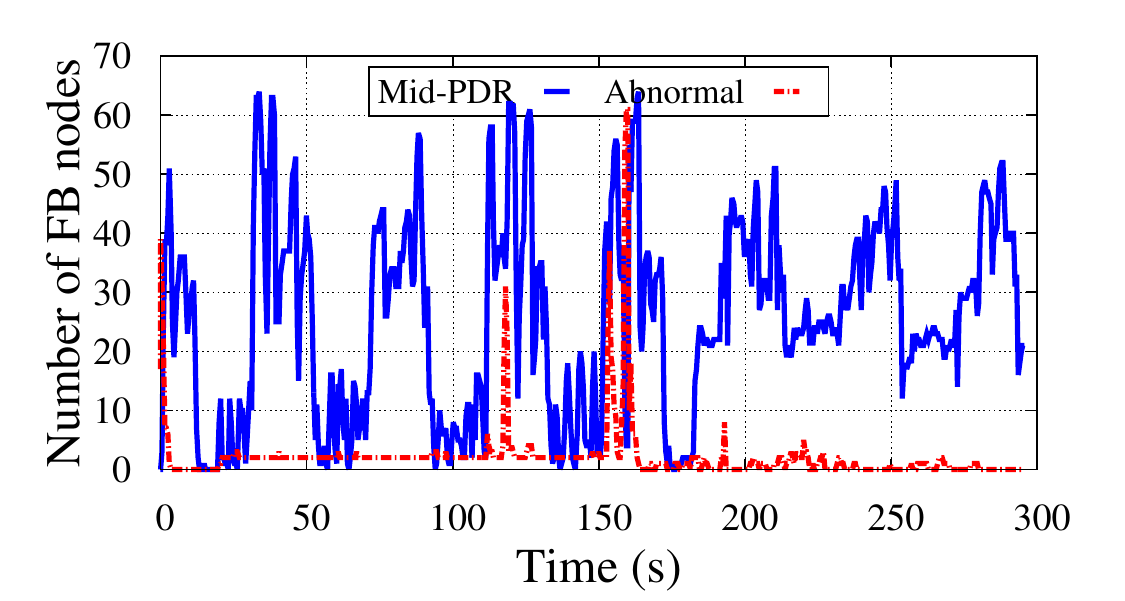}
\label{FIG:nodes_interferencenomobility} 
}%
\subfigure[]{
\includegraphics[trim=5mm 0mm 5mm 0mm, width=0.3\textwidth]{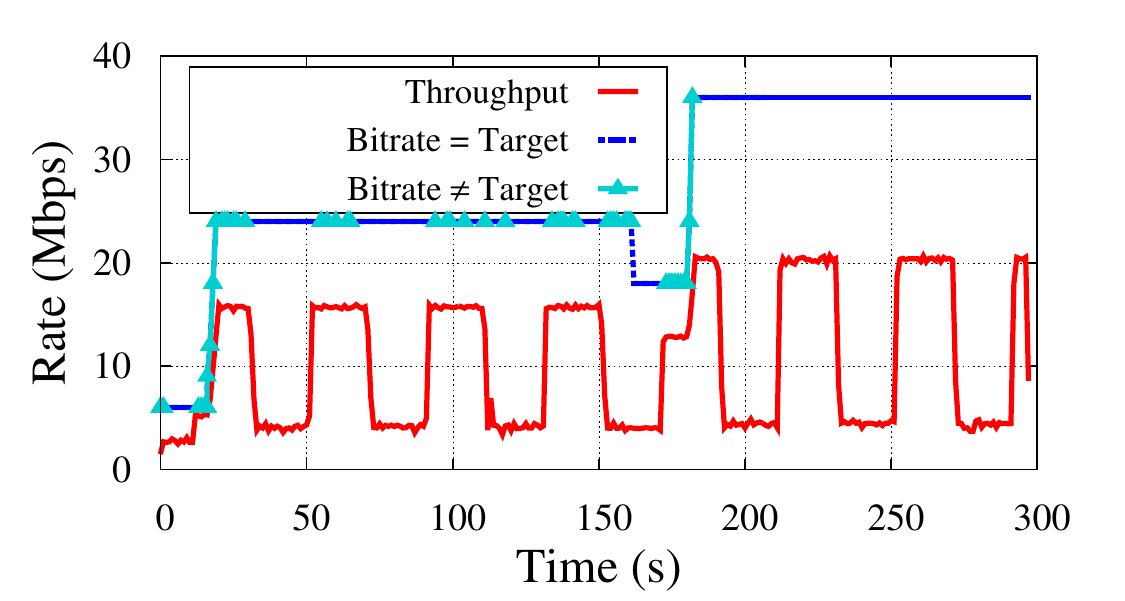}
\label{FIG:thput_interferencenomobility} 
}%
\subfigure[]{
\includegraphics[trim=5mm 0mm 5mm 0mm, width=0.3\textwidth]{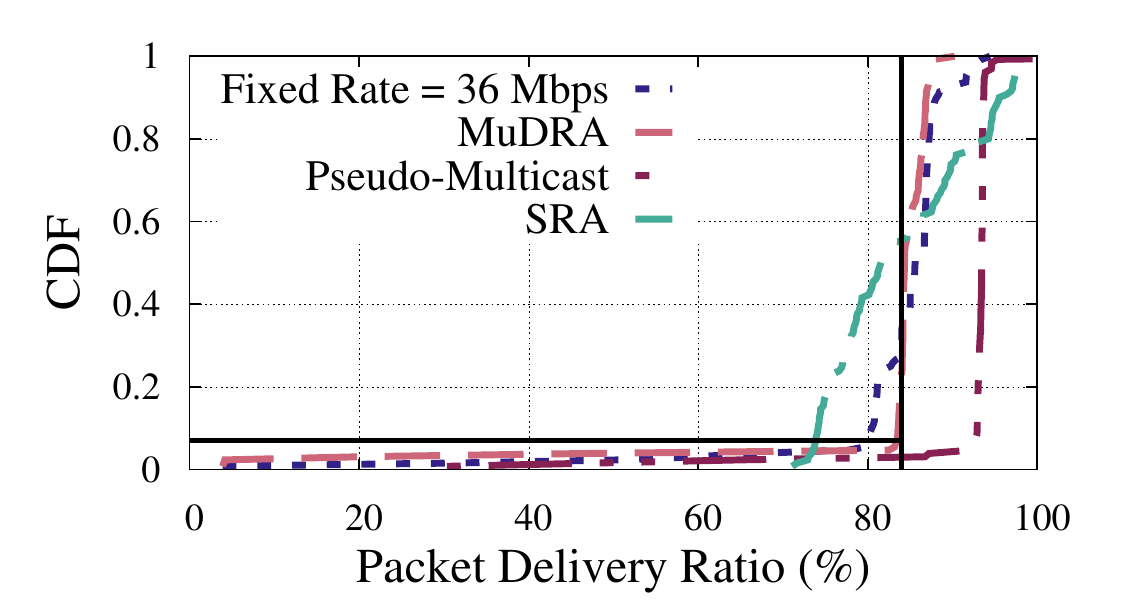}
\label{FIG:RAComparison_noise} 
}
\caption{Performance of \MUDRA with 155 nodes where an interfering AP transmits on/off traffic: (a) Mid-PDR and abnormal feedback nodes, (b) Multicast rate and throughput, (c) CDF for PDR distribution with interference for fixed rate, \MUDRANB, pseudo-multicast, SRA.} 
\vspace*{-0.4cm}
\label{FIG:interference}
\end{figure*}

We conducted 3 experiments consisting of 155 nodes (initially, 77 nodes in on state). Fig.~\ref{FIG:mobilityage} shows the impact of $p$ on the distribution of time duration that the nodes remain as FB nodes. Higher values of $p$ imply higher mobility and lead to shorter periods for which nodes serve as FB nodes.
The average number of changes in FB nodes per second is $2$, $5$, and $10$ for $p$ equal to $0$, $0.2$, and $0.9$, respectively. 
Even with these changes, the average control overhead is very low ($35$Kbps) and is not affected by the degree of mobility. 
Fig.~\ref{FIG:ratep=20} shows one instance of the RA process with $p=0.2$. We see that \MUDRA can adapt to the changing target rate at times $10$, $30$, and $255$s.
Fig.~\ref{FIG:mobilityrate} shows the percentage of data sent at different rates for several values of $p$ averaged over 3 different experiment runs. \MUDRA achieves a similar rate distribution for all values of $p$.
{\em Our experiments show that \MUDRA can achieve the target rate, maintain stability, and adds low overhead, even under severe mobility.}

\subsection{Impact of External Interference}
\label{SSC:Interference}

\ifJRNL
\begin{figure}[t]
\centering
\includegraphics[trim=5mm 0mm 10mm 0mm, width=0.3\textwidth]{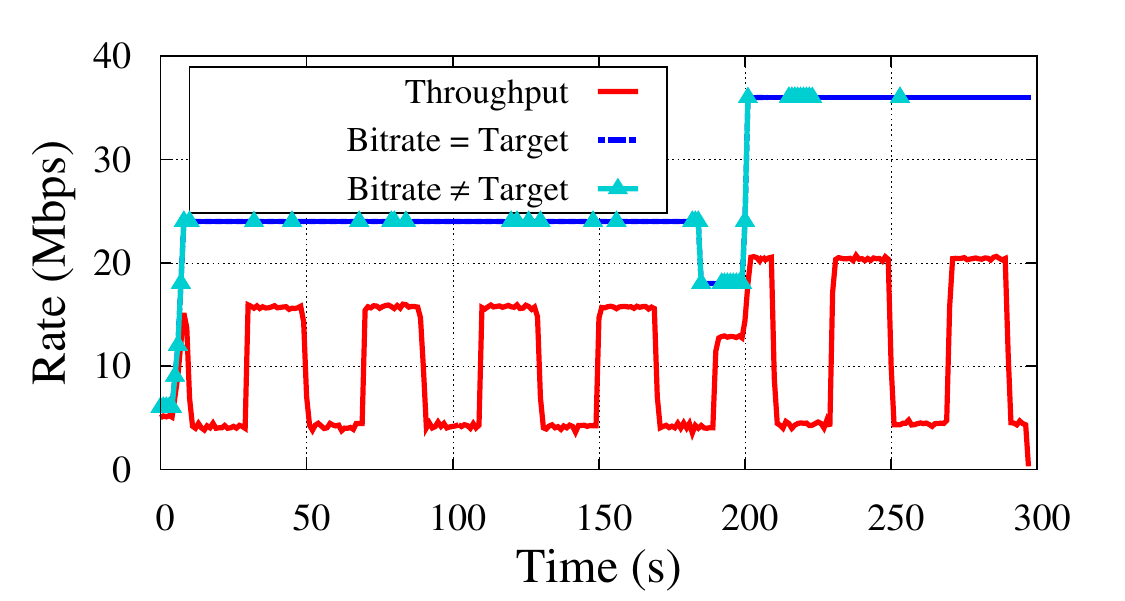}
\caption{Multicast throughput with node 1-8 transmitting interfering on/off packet stream with node mobility.}
\label{FIG:thput_interferencemobility}
\vspace{-0.25cm}
\end{figure}
\fi

We envision that \MUDRA will be deployed in environments where the wireless infrastructure is centrally controlled. However, in-channel interference can arise from mobile nodes and other wireless transmissions. 
In addition to the uncontrolled interference spikes on ORBIT,
we evaluate the impact of interference from a nearby node which transmits at the same channel as the multicast AP. We consider a scenario with two nodes near the center of the grid that exchange 
unicast traffic
at a fixed rate of $6$Mbps in a periodic on/off pattern with on and off periods $20$s each. 
The transmission power of the interfering nodes is also $0$dBm.
This helps us evaluate the performance in the worst case scenario of continuous interference and study the dynamics of changing interference. 

Fig.~\ref{FIG:nodes_interferencenomobility} shows the mid-PDR and abnormal nodes and Fig.~\ref{FIG:thput_interferencenomobility} shows the rate and throughput for one experiment with $155$ nodes.
The number of mid-PDR nodes increases during the interference periods, due to losses from collisions. \MUDRA converges to the target rate of $24$Mbps.
{\em Notice during interference periods, \MUDRA satisfied the target-condition and that using the stability preserving method, \MUDRA manages to preserve a stable rate.}  
The average throughput of different schemes with on/off background traffic for 3 experiments of $300$s each is in Table~\ref{TB:RAThroughput}. Pseudo-multicast achieves half while SRA has a third of the throughput of \MUDRANB. The fixed rate scheme achieves similar throughput as \MUDRANB.

The PDR distribution of nodes is in Fig.~\ref{FIG:RAComparison_noise}. \emph{\MUDRA satisfies QoS requirements while maintaining high throughput}. Pseudo-multicast scheme has $90\%$ nodes with PDR more than $90\%$ since it makes backoff decisions from unicast ACKs. SRA yields $55\%$ nodes with PDR less than $85\%$ as it transmits at low rates. The fixed rate scheme yields $30\%$ nodes with PDR less than $85\%$. The fixed rate scheme performs better than SRA since it maintains a higher rate.
\ifJRNL
We also investigate the combined impact of {\em both interference and mobility}, where every $6$s, the probability of a node switching on/off is $p=0.2$. Fig.~\ref{FIG:thput_interferencemobility} shows the rate and throughput for this case. Similar to results in Section~\ref{SSC:Mobility}, {\em the performance of the system is not affected by node mobility.}
\fi

\subsection{Video multicast}
\label{SSC:video}

We demonstrate the feasibility of using \MUDRA for streaming video. 
The video is segmented with segment durations equal to the period of rate changes ($1$s) and each segment is encoded at several rates in H.264 format.
For each time period, the key (I) frames are transmitted reliably at the lowest rate $6$Mbps (note that transmitting the key frames can be achieved with $100\%$ reliability even at $12$Mbps on the testbed). The non-key (B and P) frames are transmitted at the rate set by \MUDRA. 

\ifJRNL
Let the multicast rate for current time period be $R$,the expected data throughput at this rate be $\hat{D}_R$, and the estimated throughput at the minimum rate be $\hat{D}_{min}$. Let $f_k$ be the fraction of key frame data and $f_{nk}$ be the fraction of non-key frame data. The video server has to determine the video rate $V_R$ at each time $t$. 
Let the fraction of transmission time for key frames $T_k = \frac{V_R \cdot f_k}{\hat{D}_{min}}$ and fraction of transmission time for non-key frames $T_{nk} = \frac{V_R \cdot f_{nk}}{\hat{D}_R}$. 
We know that
\begin{eqnarray*}
t_k + t_{nk} = 1
\end{eqnarray*}

The video rate can be calculated by solving linear equations $V_R = \frac{\hat{D}_{min} \cdot \hat{D}_R}{\hat{D}_{min} \cdot f_{nk} + \hat{D}_R \cdot f_k}$.
In environments where estimates of throughput are inaccurate due to interference, techniques such as in \cite{video-conext} can be utilized.
\fi

\noindent
\textbf{Experimental Results:}
We use raw videos from an online dataset \cite{videodata} and encode the videos with H.264 standard. In our data sets, $f_k$ is $15-20\%$. For \MUDRA with throughput $19$Mbps and FEC correction of $15\%$, we can support a video rate of $13-15$ Mbps, which is sufficient for 3 or 4 HD streams (each $4$Mbps) on mobile devices. 
For each node, we generated the video streams offline by mapping the video frames to the detailed packet traces collected on ORBIT from an RA experiment.
In our experiments, we only considered a single video stream of rate $V_R$. For a fair comparison, the 	I frames were transmitted at $6$Mbps for all schemes.
We measured the PSNR of the video at each node and classified the PSNR in 5 categories based on visual perception.

\begin{figure}[t]
\centering
\includegraphics[trim=5mm 0mm 10mm 0mm, width=0.3\textwidth]{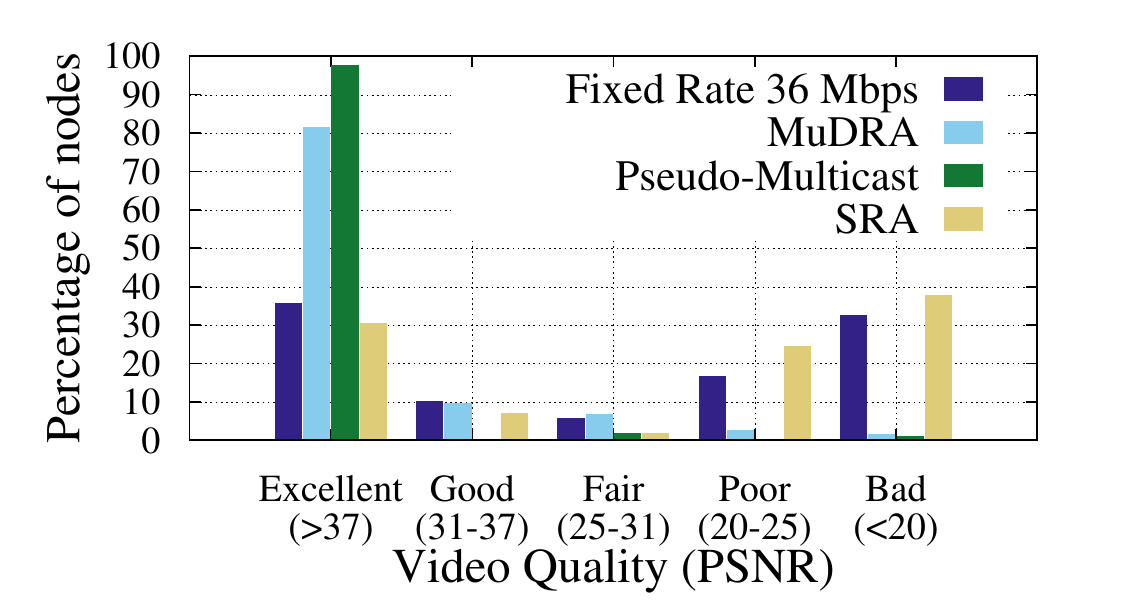}
\caption{Distribution of video quality and PSNR (in brackets) measured at 160 nodes for different multicast schemes.}
\label{FIG:PSNR}
\vspace*{-0.2cm}
\end{figure}

Fig.~\ref{FIG:PSNR} shows the video quality and PSNR ranges at the nodes for 3 experiments each of $300$s and with $150-160$ nodes. With \MUDRANB, more than $90\%$ of the nodes achieve excellent or good quality, $5\%$ achieve fair quality, and less than $5\%$ get poor or bad quality. While the pseudo-multicast scheme results in almost all nodes obtaining excellent quality, the video throughput for this scheme is significantly lower ($8$Mbps). SRA and the fixed rate schemes have more than $50\%$ nodes with poor or bad video quality.

\section{Conclusion and Future Work}
\label{SC:Conclusion}

We designed a novel multicast rate adaptation algorithm (\emph{MuDRA}) that provides high throughput while satisfying SLA requirements. \emph{MuDRA's} performance on the ORBIT testbed with hundreds of nodes shows that it can reliably support applications such as large scale multimedia content delivery. 
In future work, we will refine \MUDRA by distinguishing between losses due to channel conditions and collisions. 

\section{Acknowledgements}
\label{SC:Ack}

The authors express their gratitude to Ivan Seskar from WINLAB (Rutgers University) for his support in conducting experiments on the ORBIT testbed and several useful technical discussions.
This work was supported in part by NSF grant CNS-10-54856, CIAN NSF ERC under grant EEC-0812072, and the People Programme (Marie Curie Actions) of the European Union's Seventh Framework Programme (FP7/20072013) under REA grant
agreement no. [PIIF-GA-2013-629740].11.

\footnotesize 
\bibliographystyle{IEEEtran}
\bibliography{RA_references}

\begin{thebibliography}{10}
\providecommand{\url}[1]{#1}
\csname url@samestyle\endcsname
\providecommand{\newblock}{\relax}
\providecommand{\bibinfo}[2]{#2}
\providecommand{\BIBentrySTDinterwordspacing}{\spaceskip=0pt\relax}
\providecommand{\BIBentryALTinterwordstretchfactor}{4}
\providecommand{\BIBentryALTinterwordspacing}{\spaceskip=\fontdimen2\font plus
\BIBentryALTinterwordstretchfactor\fontdimen3\font minus
  \fontdimen4\font\relax}
\providecommand{\BIBforeignlanguage}[2]{{%
\expandafter\ifx\csname l@#1\endcsname\relax
\typeout{** WARNING: IEEEtran.bst: No hyphenation pattern has been}%
\typeout{** loaded for the language `#1'. Using the pattern for}%
\typeout{** the default language instead.}%
\else
\language=\csname l@#1\endcsname
\fi
#2}}
\providecommand{\BIBdecl}{\relax}
\BIBdecl

\bibitem{TYY10}
Y.~Tanigawa, K.~Yasukawa, and K.~Yamaoka, ``Transparent unicast translation to
  improve quality of multicast over wireless {LAN},'' in \emph{IEEE CCNC'10},
  2010.

\bibitem{CiscoStadium}
\BIBentryALTinterwordspacing
``Cisco, white-paper, {Cisco} connected stadium {W}i-{F}i solution,'' 2011.
  [Online]. Available:
  \url{http://www.cisco.com/web/strategy/docs/sports/c78-675063_dSheet.pdf}
\BIBentrySTDinterwordspacing

\bibitem{YinzCam}
``Yinzcam,'' http://www.yinzcam.com/.

\bibitem{pelechrinis-wintech10}
K.~Pelechrinis, T.~Salonidis, H.~Lundgren, and N.~Vaidya, ``Experimental
  characterization of 802.11n link quality at high rates,'' in \emph{ACM
  WiNTECH'10}, 2010.

\bibitem{hajlaoui-wintech12}
N.~Hajlaoui and I.~Jabri, ``On the performance of {IEEE} 802.11n protocol,'' in
  \emph{ACM WiNTECH'12}, 2012.

\bibitem{mcastproblem}
C.~Perkins and M.~McBride, ``Multicast {WiFi} problem statement,'' Working
  Draft, IETF Internet-Draft, 2015,
  \url{http://www.ietf.org/internet-drafts/draft-mcbride-mboned-wifi-mcast-problem-statement-00.txt}.

\bibitem{WMLLM09}
M.~Wu, S.~Makharia, H.~Liu, D.~Li, and S.~Mathur, ``{IPTV} multicast over
  wireless {LAN} using merged hybrid {ARQ} with staggered adaptive {FEC},''
  \emph{IEEE Trans. Broadcast.}, vol.~55, no.~2, pp. 363 --374, 2009.

\bibitem{CKM+09}
R.~Chandra, S.~Karanth, T.~Moscibroda, V.~Navda, J.~Padhye, R.~Ramjee, and
  L.~Ravindranath, ``{DirCast:} a practical and efficient {Wi-Fi} multicast
  system,'' in \emph{IEEE ICNP'09}, 2009.

\bibitem{Medusa}
S.~Sen, N.~K. Madabhushi, and S.~Banerjee, ``Scalable {WiFi} media delivery
  through adaptive broadcasts,'' in \emph{USENIX NSDI'10}, 2010.

\bibitem{amuse}
Y.~Bejerano, J.~Ferragut, K.~Guo, V.~Gupta, C.~Gutterman, T.~Nandagopal, and
  G.~Zussman, ``Scalable {WiFi} multicast services for very large groups,'' in
  \emph{IEEE ICNP'13}, 2013.

\bibitem{Papagiannaki2006HomeWiFi}
K.~Papagiannaki, M.~Yarvis, and W.~S. Conner, ``Experimental characterization
  of home wireless networks and design implications,'' in \emph{IEEE
  INFOCOM'06}, 2006.

\bibitem{RF:Rappaport02:wireless_com}
T.~S. Rappaport, \emph{Wireless Communication Principle and Practice, 2nd
  edition}.\hskip 1em plus 0.5em minus 0.4em\relax Prentice Hall, 2002.

\bibitem{Aguayo04interference-locality}
D.~Aguayo, J.~Bicket, S.~Biswas, G.~Judd, and R.~Morris, ``Link-level
  measurements from an 802.11b mesh network,'' \emph{ACM SIGCOMM'04}, 2004.

\bibitem{Nguyen2011RAmultiAntSys}
D.~Nguyen and J.~Garcia-Luna-Aceves, ``A practical approach to rate adaptation
  for multi-antenna systems,'' in \emph{IEEE ICNP'11}, 2011.

\bibitem{Combes2014RAreport}
R.~Combes, A.~Proutiere, D.~Yun, J.~Ok, and Y.~Yi, ``Optimal rate sampling in
  802.11 systems,'' \emph{arXiv preprint}, 2013.

\bibitem{multicastsurvey}
J.~Vella and S.~Zammit, ``A survey of multicasting over wireless access
  networks,'' \emph{IEEE Commun. Surveys Tuts.}, vol.~15, no.~2, pp. 718--753,
  2013.

\bibitem{Kamerman97WaveLan}
A.~Kamerman and L.~Montebani, ``{WaveLAN}-ii: a high-performance wireless {LAN}
  for the unlicensed band,'' \emph{Bell Labs technical journal}, vol.~2, no.~3,
  p. 118–133, 1997.

\bibitem{Lacage2004RA}
M.~Lacage, M.~Manshaei, and T.~Turletti, ``{IEEE} 802.11 rate adaptation: a
  practical approach,'' in \emph{ACM MSWiM'04}, 2004.

\bibitem{Bicket2005RA}
J.~Bicket, ``Bit-rate selection in wireless networks,'' \emph{PhD thesis, MIT},
  2005.

\bibitem{Pang2005LDARF}
Q.~Pang, V.~Leung, and S.~Liew, ``A rate adaptation algorithm for {IEEE} 802.11
  {WLANs} based on {MAC}-layer loss differentiation,'' in \emph{IEEE
  BroadNets'06}, 2006.

\bibitem{Wong2006RRAA}
S.~Wong, H.~Yang, S.~Lu, and V.~Bharghavan, ``Robust rate adaptation for 802.11
  wireless networks,'' in \emph{ACM MOBICOM'06}, 2006.

\bibitem{Kim2006CARA}
J.~Kim, S.~Kim, S.~Choi, and D.~Qiao, ``{CARA:} collision-aware rate adaptation
  for {IEEE 802.11 WLANs},'' in \emph{IEEE INFOCOM'06}, 2006.

\bibitem{Radunovic11WhiteSpaces}
B.~Radunovic, A.~Proutiere, D.~Gunawardena, and P.~Key, ``Dynamic channel, rate
  selection and scheduling for white spaces,'' in \emph{ACM CONEXT'11}, 2011.

\bibitem{Combes2014RA}
R.~Combes, A.~Proutiere, D.~Yun, J.~Ok, and Y.~Yi, ``Optimal rate sampling in
  802.11 systems,'' in \emph{IEEE INFOCOM'14}, 2014.

\bibitem{Holland2001SNR-RA}
G.~Holland, N.~Vaidya, and P.~Bahl, ``A rate-adaptive mac protocol for
  multi-hop wireless networks,'' in \emph{ACM MOBICOM'01}, 2001.

\bibitem{Rayanchu2008COLLIE}
S.~Rayanchu, A.~Mishra, D.~Agrawal, S.~Saha, and S.~Banerjee, ``Diagnosing
  wireless packet losses in 802.11: Separating collision from weak signal,'' in
  \emph{IEEE INFOCOM'08}, 2008.

\bibitem{Judd2008CHARM}
G.~Judd, X.~Wang, and P.~Steenkiste, ``Efficient channel-aware rate adaptation
  in dynamic environments,'' in \emph{ACM MobiSys'08}, 2008.

\bibitem{Balakrishnan2009SoftRate}
M.~Vutukuru, H.~Balakrishnan, and K.~Jamieson, ``Cross-layer wireless bit rate
  adaptation,'' in \emph{ACM SIGCOMM'09}, 2009.

\bibitem{Katabi2009FARA}
H.~Rahul, F.~Edalat, D.~Katabi, and C.~Sodinii, ``Frequency-aware rate
  adaptation and {MAC} protocols,'' in \emph{ACM MOBICOM'09}, 2009.

\bibitem{Crepaldi2012CSISF}
R.~Crepaldi, J.~Lee, R.~Etkin, S.-J. Lee, and R.~Kravets, ``{CSI-SF}:
  Estimating wireless channel state using {CSI} sampling and fusion,'' in
  \emph{IEEE INFOCOM'12}, 2012.

\bibitem{Belding2013ARAMIS}
L.~Deek, E.~Garcia-Villegas, E.~Belding, S.-J. Lee, and K.~Almeroth, ``Joint
  rate and channel width adaptation in 802.11 {MIMO} wireless networks,'' in
  \emph{IEEE SECO3'13}, 2013.

\bibitem{KK01}
J.~K. Kuri and S.~Kumar, ``Reliable multicast in multi-access wireless
  {LANs},'' \emph{ACM/Kluwer Wirel. Netw.}, vol.~7, pp. 359--369, 2001.

\bibitem{SHAL02}
M.-T. Sun, L.~Huang, A.~Arora, and T.-H. Lai, ``Reliable {MAC} layer multicast
  in {IEEE} 802.11 wireless networks,'' in \emph{IEEE ICPP'02}, 2002.

\bibitem{VCOST07}
J.~Villalon, P.~Cuenca, L.~Orozco-Barbosa, Y.~Seok, and T.~Turletti,
  ``Cross-layer architecture for adaptive video multicast streaming over
  multirate wireless {LAN}s,'' \emph{IEEE J. Sel. Areas Commun.}, vol.~25,
  no.~4, pp. 699 --711, 2007.

\bibitem{CSKC10}
N.~Choi, Y.~Seok, T.~Kwon, and Y.~Choi, ``Leader-based multicast service in
  {IEEE} 802.11v networks,'' in \emph{IEEE CCNC'10}, 2010.

\bibitem{LH08:BLBP}
Z.~Li and T.~Herfet, ``{BLBP:} a beacon-driven leader based protocol for {MAC}
  layer multicast error control in wireless {LAN}s,'' in \emph{IEEE WiCOM'08},
  2008.

\bibitem{SR09}
V.~Srinivas and L.~Ruan, ``An efficient reliable multicast protocol for
  802.11-based wireless {LAN}s,'' in \emph{IEEE WoWMoM'09}, 2009.

\bibitem{WWWZY09}
X.~Wang, L.~Wang, Y.~Wang, Y.~Zhang, and A.~Yamada, ``Supporting {MAC} layer
  multicast in {IEEE} 802.11n: Issues and solutions,'' in \emph{IEEE WCNC'09},
  2009.

\bibitem{PHKSJ08}
E.~Park, S.~Han, H.~Kim, K.~Son, and L.~Jing, ``Efficient multicast video
  streaming for {IPTV} service over {WLAN} using {CC-FEC},'' in \emph{IEEE
  ICICSE'08}, 2008.

\bibitem{AKWP10}
O.~Alay, T.~Korakis, Y.~Wang, and S.~Panwar, ``Dynamic rate and {FEC}
  adaptation for video multicast in multi-rate wireless networks,'' \emph{ACM
  Mobile Netw. and Appl.}, vol.~15, no.~3, pp. 425--434, 2010.

\bibitem{CCLKT12}
H.-T. Chiao, S.-Y. Chang, K.-M. Li, Y.-T. Kuo, and M.-C. Tseng, ``{WiFi}
  multicast streaming using {AL-FEC} inside the trains of high-speed rails,''
  in \emph{IEEE BMSB'12}, 2012.

\bibitem{SC03}
Y.~Seok and Y.~Choi, ``Efficient multicast supporting in multi-rate wireless
  local area networks,'' in \emph{IEEE ICOIN'03}, 2003.

\bibitem{BSS06}
A.~Basalamah, H.~Sugimoto, and T.~Sato, ``Rate adaptive reliable multicast
  {MAC} protocol for {WLANs},'' in \emph{IEEE VTC'06}, 2006.

\bibitem{LKS12}
W.-S. Lim, D.-W. Kim, and Y.-J. Suh, ``Design of efficient multicast protocol
  for {IEEE} 802.11n {WLAN}s and cross-layer optimization for scalable video
  streaming,'' \emph{IEEE Trans. Mobile Comput.}, vol.~11, no.~5, pp. 780
  --792, 2012.

\bibitem{WWWG08}
X.~Wang, L.~Wang, and D.~Wang, Y.and~Gu, ``Reliable multicast mechanism in
  {WLAN} with extended implicit {MAC} acknowledgment,'' in \emph{IEEE VTC'08},
  2008.

\bibitem{SL09}
V.~Srinivas and L.~Ruan, ``An efficient reliable multicast protocol for
  802.11-based wireless {LAN}s,'' in \emph{IEEE WoWMoM'09}, 2009.

\bibitem{FWYL10}
Z.~Feng, G.~Wen, C.~Yin, and H.~Liu, ``Video stream groupcast optimization in
  {WLAN},'' in \emph{IEEE ITA'10}, 2010.

\bibitem{802.11aa}
``{IEEE} draft standard for information technology telecommunications and
  information exchange between systems local and metropolitan area networks -
  specific requirements, part 11: Wireless {LAN} medium access control ({MAC})
  and physical layer ({PHY}) specifications - amendment: {MAC} enhancements for
  robust audio video streaming,'' IEEE, July 2011.

\bibitem{PJYK10}
Y.~Park, C.~Jo, S.~Yun, and H.~Kim, ``Multi-room {IPTV} delivery through
  pseudo-broadcast over {IEEE} 802.11 links,'' in \emph{IEEE VTC'10}, 2010.

\bibitem{LH08:HLBP}
Z.~Li and T.~Herfet, ``{HLBP}: a hybrid leader based protocol for {MAC} layer
  multicast error control in wireless {LAN}s,'' in \emph{IEEE GLOBECOM'08},
  2008.

\bibitem{amuseGree}
Y.~Bejerano, J.~Ferragut, K.~Guo, V.~Gupta, C.~Gutterman, T.~Nandagopal, and
  G.~Zussman, ``Experimental evaluation of a scalable {WiFi} multicast scheme
  in the {ORBIT} testbed,'' in \emph{3rd GENI Research and Educational
  Experiment Workshop (GREE)}, 2014.

\bibitem{SSFNB09}
V.~Sgardoni, M.~Sarafianou, P.~Ferre, A.~Nix, and D.~Bull, ``Robust video
  broadcasting over 802.11a/g in time-correlated fading channels,'' \emph{IEEE
  Trans. Consum. Electron.}, vol.~55, no.~1, pp. 69--76, 2009.

\bibitem{LSHC11}
K.~Lin, W.~Shen, C.~Hsu, and C.~Chou, ``Quality-differentiated video multicast
  in multi-rate wireless networks,'' \emph{IEEE Trans. Mobile Comput.},
  vol.~12, no.~1, pp. 21--34, January 2013.

\bibitem{ORBIT}
``{ORBIT} testbed,'' http://orbit-lab.org/.

\bibitem{Nybomand07FEC}
K.~N.~D. Vukobratovic, ``A survey on application layer forward error correction
  codes for {IP} datacasting in {DVB-H},'' in \emph{3rd COST 2100 MCM}, 2007.

\bibitem{Ababneh14FEC}
J.~Ababneh and O.~Almomani, ``Survey of error correction mechanisms for video
  streaming over the internet,'' in \emph{IJACSA}, vol.~5, 2014, pp. 155--161.

\bibitem{minstrel}
``Minstrel,''
  \url{https://wireless.wiki.kernel.org/en/developers/documentation/mac80211/ratecontrol/minstrel}.

\bibitem{cranley2006user}
N.~Cranley, P.~Perry, and L.~Murphy, ``User perception of adapting video
  quality,'' \emph{Int. J. Hum. Comput. St.}, vol.~64, no.~8, pp. 637--647,
  2006.

\bibitem{balachandran2012quest}
A.~Balachandran, V.~Sekar, A.~Akella, S.~Seshan, I.~Stoica, and H.~Zhang, ``A
  quest for an internet video quality-of-experience metric,'' in \emph{ACM
  HotNets'12}, 2012.

\bibitem{Reis2006interference}
C.~Reis, R.~Mahajan, M.~Rodrig, D.~Wetherall, and J.~Zahorjan,
  ``Measurement-based models of delivery and interference in static wireless
  networks,'' in \emph{ACM SIGCOMM'06}, 2006.

\bibitem{Halperin2010wifi}
D.~Halperin, W.~Hu, A.~Sheth, and D.~Wetherall, ``Predictable 802.11 packet
  delivery from wireless channel measurements,'' in \emph{ACM SIGCOMM'10},
  2010.

\bibitem{souryal}
M.~R. Souryal, L.~Klein-Berndt, L.~E. Miller, and N.~Moayeri, ``Link assessment
  in an indoor 802.11 network,'' in \emph{IEEE WCNC'06}, 2006.

\bibitem{video-conext}
G.~Tian and Y.~Liu, ``Towards agile and smooth video adaptation in dynamic http
  streaming,'' in \emph{ACM CONEXT'12}, 2012.

\bibitem{videodata}
``Video dataset,'' \url{https://media.xiph.org/video/derf/}.

\end{thebibliography}

\end{document}